\documentclass[structabstract]{aa}  
\usepackage{graphicx}
\usepackage{txfonts}
\usepackage[round]{natbib}
\usepackage{lscape}
\begin{document}
   \title{APEX observations of supernova remnants}

   \subtitle{I. Non-stationary MHD-shocks in W44\thanks{The velocity-integrated CO maps shown in Fig. \ref{Fig:3} and \ref{Fig:4} are available as FITS files at the CDS via anonymous ftp to cdsarc.u-strasbg.fr (130.79.128.5) or via http://cdsweb.u-strasbg.fr/cgi-bin/qcat?J/A+A/ .}}

   \author{ S. Anderl
          \inst{1,2,3}
          \and
          A. Gusdorf\inst{4}
          \and
          R. G\"usten\inst{5}
          }

   \institute{Argelander Institut f\"ur Astronomie, Universit\"at Bonn, Auf dem H\"ugel, 71, 53121 Bonn, Germany
   \and
   Univ. Grenoble Alpes, IPAG, F-38000 Grenoble, France
   \and
CNRS, IPAG, F-38000 Grenoble, France\\   \email{sibylle.anderl@obs.ujf-grenoble.fr}
   \and
   LERMA, UMR 8112 du CNRS, Observatoire de Paris, \'Ecole Normale Sup\'erieure, 24 rue Lhomond, 75231 Paris Cedex 05, France
   \and
   Max Planck Institut f\"ur Radioastronomie, 
              Auf dem H\"ugel 69, 53121 Bonn, Germany
             }

   \date{Received .................; accepted .................}

% \abstract{}{}{}{}{} 
% 5 {} token are mandatory
  \abstract
  % context heading (optional)
  % {} leave it empty if necessary  
   {When supernova blast waves interact with nearby molecular clouds, they send slower shocks into these clouds. The resulting interaction regions provide excellent environments for the use of MHD shock models to constrain the physical and chemical conditions in these regions.}
  % aims heading (mandatory)
   {The interaction of supernova remnants (SNRs) with molecular clouds gives rise to strong molecular emission in the far-IR and sub-mm wavelength regimes. The application of MHD shock models in the interpretation of this line emission can yield valuable information on the energetic and chemical impact of supernova remnants.
   }
  % methods heading (mandatory)
   {New mapping observations with the APEX telescope in $^{12}$CO (3--2), (4--3), (6--5), (7--6) and $^{13}$CO (3--2) towards two regions in the supernova remnant W44 are presented. Integrated intensities are extracted on five different positions, corresponding to local maxima of CO emission. The integrated intensities are compared to the outputs of a grid of models, which combine an MHD shock code with a radiative transfer module based on the `large velocity gradient' approximation.}
  % results heading (mandatory)
   {All extracted spectra show ambient and line-of-sight components as well as blue- and red-shifted wings indicating the presence of shocked gas.
     Basing the shock model fits only on the highest-lying transitions that unambiguously trace the shock-heated gas, we find that the observed CO line emission is compatible with non-stationary shocks and a pre-shock density of 10$^4$ cm$^{-3}$. The ages of the modelled shocks scatter between values of $\sim$1000 and $\sim$3000 years. The shock velocities in W44F are found to lie between 20 and 25 km s$^{-1}$, while in W44E fast shocks (30--35 km s$^{-1}$) as well as slower shocks ($\sim$20 km s$^{-1}$) are compatible with the observed spectral line energy diagrams. The pre-shock magnetic field strength components perpendicular to the line-of-sight in both regions have values between 100 and 200 $\mu$G. Our best-fitting models allow us to predict the full ladder of CO transitions, the shocked gas mass in one beam as well as the momentum- and energy injection.
   }
  % conclusions heading (optional), leave it empty if necessary 
   {}


   \maketitle
%
%________________________________________________________________

\section{Introduction}\label{Sec:1}

Supernova explosions strongly affect the dynamical state of the interstellar medium (ISM). These explosions represent an injection of about 10${}^{51}$~erg into the ISM, immediately creating regions of very hot and tenous gas (e.g. \citealt{Cox:2005}). The shock waves originating from these explosions disperse molecular clouds, sweep up and compress the ambient medium (\citealt{McKee:1977}). The evolution of the supernova remnant (SNR) can be described by four successive phases (\citealt{Woltjer:1972}): a free expansion phase, where the density of the ejected matter is much larger than that of the surrounding medium, a phase of adiabatic expansion, where the gas is too hot to undergo efficient radiative cooling, a radiative phase, where a dense, radiatively cooling shell of swept up gas is formed, and finally a fadeaway phase, when the shock wave has slowed down so much that it has turned into a sound wave.

If the supernova blast wave encounters ambient molecular clouds, it drives slower shocks into these clouds at a velocity that relates to the shock velocity in the intercloud medium via the density contrast between cloud- and intercloud gas (\citealt{McKee:1975}). These slow shocks are similar to shocks originating in bipolar outflows of very young stars (e.g. \citealt{Gusdorf:2011}): they strongly cool through molecular emission and can be observed in the far-IR and sub-mm wavelength regime (e.g. \citealt{Neufeld:2007,Frail:1998p16306}). Clear evidence for an interaction between a SNR and molecular clouds is typically provided by the detection of broad line wings, maser emission, highly excited far-IR CO and near- and mid-infrared H$_2$ emission (\citealt{Reach:2005p16099}). However, the modelling of these shocks differs from those associated with outflows in star forming cores because they are not irradiated by an embedded proto-star and they do not show any envelope or infall processes. The study of these interactions between SNRs and molecular clouds can yield valuable information on the supernova explosion and its impact on the surrounding medium. At the same time it can improve our understanding of the molecular clouds themselves. This information is needed for the understanding of various astrophysical questions, such as the energy balance of the ISM in galaxies, triggered star formation, and the origin and acceleration of cosmic rays.

High-$J$ CO line emission is a very good diagnostic in this context because of the high abundance of CO in the molecular ISM, its important role for the cooling of the medium, and  the fact that its rotational transitions between higher energy levels are expected to trace shock conditions (e.g. \citealt{Flower:2010,Meijerink:2013}). Here, we present new APEX CO observations towards two regions in one of the prototype interacting SNRs, W44, which has been the subject of many observational studies. We find MHD models compatible with our observations of CO emission and explore the consequences of the corresponding shock scenarios with respect to the impact of the SNR on its environment.

This paper builds on our previous study of MHD shocks based on high-$J$ CO observations towards the SNR W28 (\citealt{Gusdorf:2012p18545}), but refines and extends the methods established there. The structure of the paper is as follows. In Sect. \ref{Sec:2}, we give a review of the supernova remnant W44. In Sect. \ref{Sec:3}, we present our observations together with more detailed information on the observed regions W44E and W44F. Sect. \ref{Sec:4} provides information on the molecular environment of W44. In Sect. \ref{Sec:5}, we show the CO maps observed towards both regions, while Sect. \ref{Sec:6} focuses on the observed spectra towards the positions of our shock analysis. Our modelling approach with respect to the processing of observations and the grid of models is described in Sect. \ref{Sec:7}. In Sect. \ref{Sec:8}, the results are presented and discussed in Sect. \ref{Sec:9}. We summarize our findings in the concluding Sect. \ref{Sec:10}.

\section{The supernova remnant W44}\label{Sec:2}

   \begin{figure*}
   \centering
   \includegraphics[width=\textwidth]{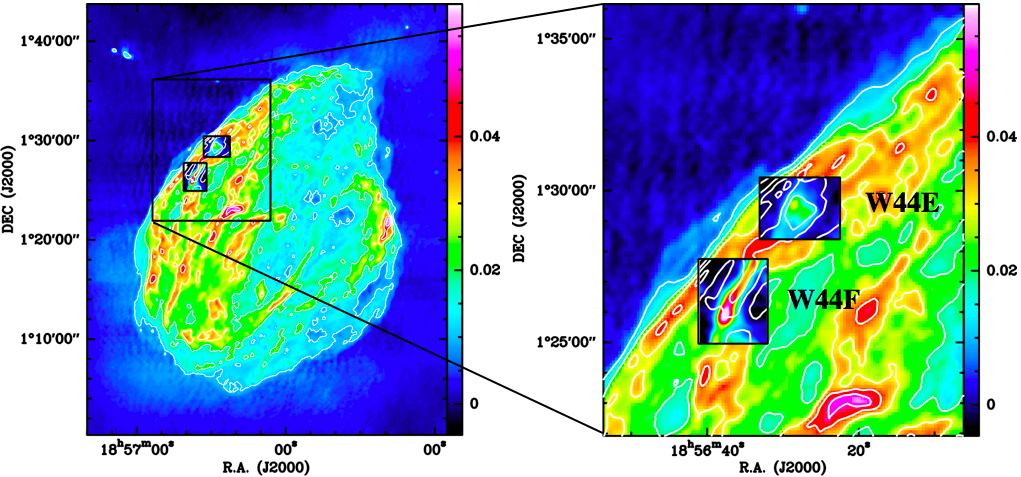}
      \caption{Location of the fields covered by our CO observations on the larger-scale radio continuum image at 1442.5 MHz, taken from Giacani et al. (1997). The wedges indicate the intensity of the continuum in Jy/beam. Left panel: entire SNR, right panel: zoom in the regions W44E and W44F. The inserts show the distribution of the CO (6--5) emission, integrated between 20 km s${}^{-1}$ and 70 km s${}^{-1}$, for which the colourscale is in the range 0 - 50 K~km s${}^{-1}$, and the contours are radio continuum in steps of 10~mJy. The size of the zoombox is $\sim$10.3 $\times$ 12.4 pc.
              }
         \label{Fig:1}
   \end{figure*}

W44 (a.k.a. G34.7--0.4, or 3C 392) is a prototype of the so-called mixed-morphology supernova remnant as described by \citet{Rho:1998p19482}. This class refers to SNRs with centrally concentrated X-ray emission and a shell-like radio morphology. The size of this semi-symmetric SNR is about 30$\arcmin$ (\citealt{Rho:1998p19482}). It is probably located in the Sagittarius arm (\citealt{Castelletti:2007p17711}) at the base of the Aquila supershell (\citealt{Maciejewski:1996p19499}) in a very obscured, complex region in the Galactic plane. Its distance has been estimated on the basis of HI 21 cm absorption measurements as $\sim$3 kpc (\citealt{Radhakrishnan:1972p16510,Caswell:1975p19765,Green89}) and confirmed as 2.9~$\pm$~0.2~kpc through molecular observations using the solar constants $R_0=7.6$ kpc and $\Omega_0 =$ 27.2~km~s${}^{-1}$ kpc${}^{-1}$ (\citealt{Castelletti:2007p17711}). In this paper we use the distance value of 3 kpc as e.g. \citet{Abdo:2010p17181}, \citet{Paron:2009p15970}, and \citet{Seta:2004p16955}. Based on this distance the SNR's size of $\sim$30$\arcmin$ corresponds to $\sim$26 pc. W44 is believed to be in a radiative phase over much of its surface due to its estimated age of  $2 \times 10^4$~years (see below) and the observed cooling radiation  (\citealt{Cox:1999p16961,Chevalier:1999p16808}).

As first reported by \citet{Wolszczan:1991p19503}, this type II SNR harbors a fast-moving 267 millisecond radio pulsar, PSR B1853+01, within the W44 radio shell, about 9$\arcmin$ south from its geometrical centre. Its progenitor is assumed to have had little influence on the parental molecular cloud on the scale of the SNR (\citealt{Reach:2005p16099}). The pulsar's spin-down age amounts to $2 \times 10^4$~years, while its distance, derived from the dispersion measure, is consistent with that of the SNR (\citealt{Taylor:1993p19504}). The pulsar wind powers a small synchrotron nebula observed in radio (\citealt{Frail:1996p19510}) and X-ray emission (\citealt{Harrus:1996p19761,Petre:2002p19762}). 

W44 was first detected as a radio source already in the late 50Õs (\citealt{Westerhout:1958p19830,Mills:1958p19834,Edge59}) and identified as a possible SNR due to its non-thermal radio spectrum (\citealt{Scheuer:1963}). The shell-type shape in radio-continuum maps appears northeast-southwest elongated, with a size of 25$\arcmin$ $\times$ 35$\arcmin$ at 1442.5~MHz (\citealt{Giacani:1997p16602}) and enhanced radio emission at the eastern portion of the SNR. A high resolution VLA radio image at 1465~MHz, observed by \citet{Jones:1993p19637}, reveals several filaments across the remnant and distortions along its eastern border. A plausible explanation for this feature could be the expansion of the SNR into a cloudy interstellar medium, as already suggested by \citet{Velusamy:1988}. 
Polarization measurements at 2.8~cm, conducted by \citet{Kundu:1972p19885}, show a high degree of polarization as large as 20\%, possibly due to a highly regular oriented magnetic field, particularly over the main eastern and northern parts of the SNR. Based on earlier measurements between 22 and 10700 MHz, and new observations at 74 and 324 MHz, \citet{Castelletti:2007p17711} derived a global integrated continuum spectral index $-0.37 \pm$0.02.

$\ion{H}{I}$ observations reveal a fast, expanding shell-like structure at $\varv{}_{\rm LSR} = 125$ and 210 km s${}^{-1}$ associated with the SNR, corresponding to an expansion velocity of the H I shell of $\varv{}_{\rm exp}$ = 150 $\pm$15 km s${}^{-1}$ (\citealt{Koo:1995p19577}). These authors estimated the H I shell as being significantly smaller than the radio continuum shell. They interpreted this double-shell structure as originating from the supernova exploding inside a pre-existing wind bubble, whose shell constitutes the observed H I structure. \citet{Shelton:1999p19589} used a hydrocode model for W44 to show that this interpretation is not compelling if the remnant evolution is taking place in a density gradient, which leads to a spread in shell formation times for different parts of the remnant's surface. While \citet{Shelton:1999p19589} assumed the ISM to be denser on the northeastern, far side of the SNR, \citet{Seta:2004p16955} expected the strongest interaction between the SNR and the ISM to occur in the front side of the SNR, while it is the redshifted side that is visible as high-velocity expanding H I shell. This conclusion was based on observed OH absorption in their component of spatially extended moderately broad emission.

  \begin{figure}
   \centering
   \includegraphics[width=9cm, height = 10cm]{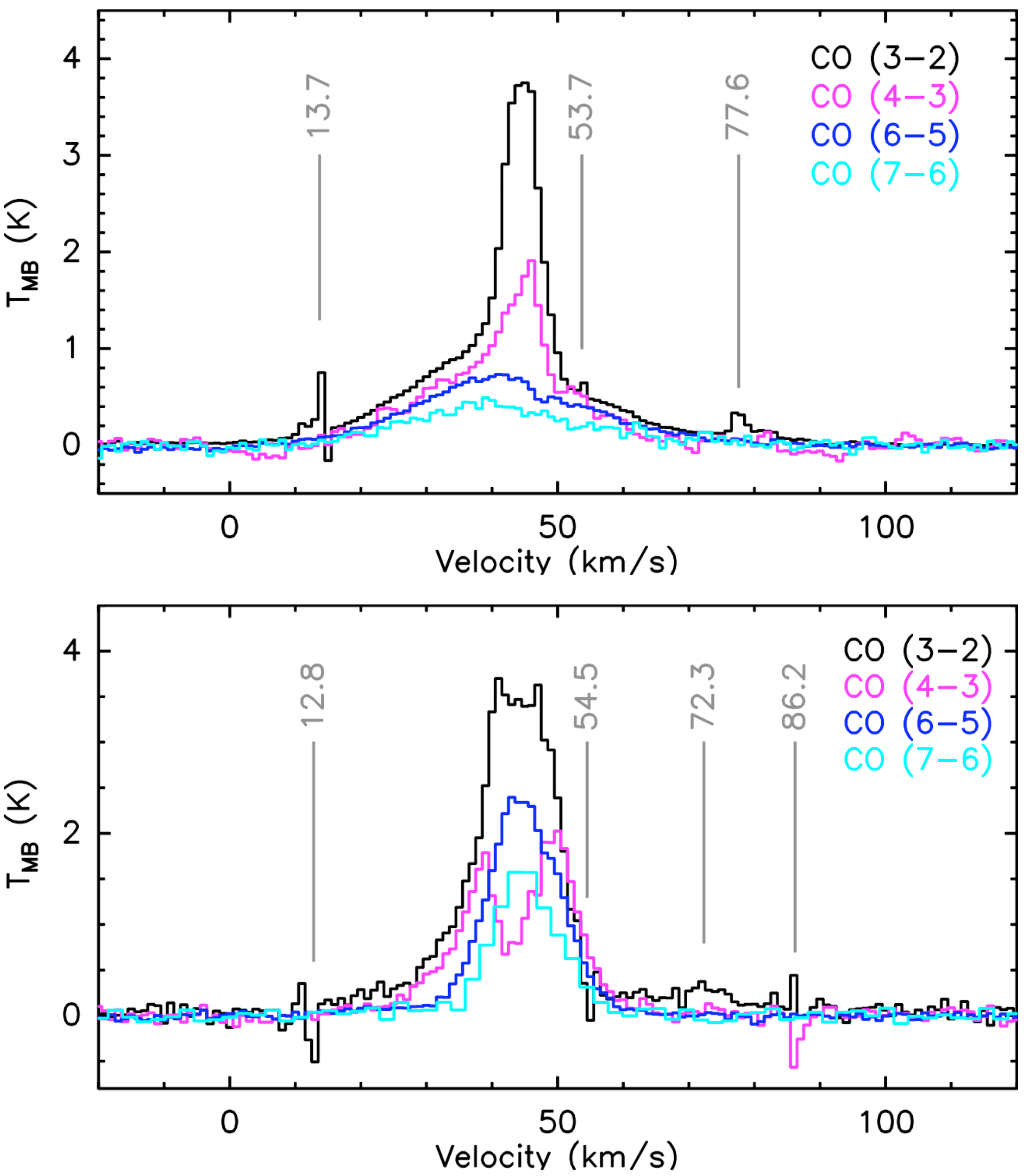}
      \caption{     CO spectra (in $T_{\rm MB}$) averaged over the W44E (top) and W44F (bottom) regions. The spectra correspond to the transitions of CO (3--2) (black), CO (4--3) (pink), CO (6--5) (blue), and CO (7--6) (light blue). The CO spectral resolutions are 2.18 km s${}^{-1}$ for CO (7--6) in W44F and 1.0 km s${}^{-1}$ for all other lines.
                     }
         \label{Fig:2}
   \end{figure}

The centrally peaked X-ray morphology was first noted in observations from the Einstein Observatory Image Proportional Counter (IPC) (\citealt{Watson:1983p19639,Smith:1985p19638}). Assuming the X-ray emission to be thermal (\citealt{Szymkowiak:1980p19924,Jones:1993p19637}), the morphology of the emission was explained by a scenario where a possible X-ray shell is unseen because it has become too cold to be detected through the intervening ISM (\citealt{Smith:1985p19638,Jones:1993p19637}). 
Observations with the ROSAT Position Sensitive Proportional Counters (PSPC) (\citealt{Rho:1994p16098}) confirmed the centrally peaked X-ray morphology and revealed a largely uniform temperature over the remnant. \citet{Rho:1994p16098} as well as \citet{Jones:1993p19637} interpreted this observation using an evaporation model with a two-phase interstellar medium structure of clump and interclump gas (\citealt{White:1991p19958}), where the material stemming from evaporating clouds increases the density of the SNR interior. More recent models stressed the important role of thermal conduction in the creation of the centrally peaked X-ray emission (\citealt{Cox:1999p16961,Kawasaki:2005p19670}). Thermal conduction levels out gradients of temperature in the hot interior plasma while pressure equilibrium then yields a higher density at the centre. Once the forward shock velocity has decreased so that the X-ray emission from the shell becomes too soft to pass through the ISM, the centrally brightened X-ray emission appears. 

Probably due to the interaction between the SNR and a molecular cloud, the eastern limb shows a lack of X-ray emission within the radio shell, while weaker diffuse X-ray emission extends up to the northern radio boundary of W44 (\citealt{Giacani:1997p16602}). The existence of a non-thermal X-ray component was shown by \citet{Harrus:1996p19761} when they detected a hard X-ray source coincident with the position of the pulsar region interpreted as an X-ray synchrotron nebula.

The first detection of optical filaments in H$\alpha$ and [S II] images of W44, seen in the north and southeast parts, was reported by \citet{Rho:1994p16098}. These filaments are mainly confined within the X-ray emitting region and are absent in the eastern region. Along the northwest border of the remnant there is excellent correlation between optical and radio emission at 1.4 GHz, suggesting that radiative cooling of the shocked gas immediately behind the shock front, where enhanced magnetic fields are present, is the origin of the optical radiation (\citealt{Giacani:1997p16602}).

The association of molecular gas with W44 was already reported by \citet{Dickel:1976p16135}, who attributed the flattened shell structure of the SNR to an encounter with a dense molecular cloud. \citet{Wootten:1977p16116} observed a broadening of line widths and an intensification of line strength in CO (1--0), which he explained by the heating and compression of the ISM by the SNR. \citet{Denoyer:1983p16515} questioned the shock processing of the adjacent molecular cloud, referring to an absence of chemical shock characteristics and the possible explanation of broadened lines by overlapping line components along the line-of-sight. However, subsequent observations of emission from shock-tracing molecules and fine-structure line emission (\citealt{Seta:1998p16736,Reach:1996p16617,Reach:2000p16359,Reach:2005p16099,Neufeld:2007p15887,Yuan:2011p15680}) as well as OH maser emission (\citealt{Claussen:1997p16194,Hoffman:2005p16236}) have removed doubts about the existence of an interaction with a molecular cloud.

Finally, W44 is an important site for the study of cosmic-ray production, acceleration and propagation based on the observations of $\gamma$-rays stemming from the interaction of cosmic-rays and the interstellar medium. These interactions include high-energy electron bremsstrahlung and inverse Compton scattering of the leptonic cosmic-ray component as well as proton-proton collisions of the hadronic component creating neutral pions, whose decay generates $\gamma$-rays. A first detection in the GeV regime was achieved with the EGRET instrument aboard the Compton Gamma Ray Observatory, although an association with W44 was not clear (\citealt{Esposito:1996p19685}). The Fermi Large Area Telescope (LAT) detected a source morphology corresponding to the SNR radio-shell at energies above 200 MeV (\citealt{Abdo:2010p17181}). Modelling the observed emission, these authors concluded that neutral pion decays are plausibly responsible for the emission, although a bremsstrahlung scenario could not be ruled out completely. \citet{Giuliani:2011p19731} subsequently excluded leptonic emission as the main contribution based on observations at lower energies (30MeV -- 50 GeV) using the AGILE instrument. They reported direct evidence for pion emission based on a steep decline of the spectral energy distribution below 1 GeV. \citet{Ackermann:2013p19994} presented observations in the sub-GeV part of the $\gamma$-ray spectrum, obtained within four years with the Fermi LAT, from the compact regions delineated by the radio continuum emission. Based on the observation of the characteristic pion-decay  feature in their spectrum  (``pion-decay bump'') they could finally provide direct evidence for the acceleration of cosmic-ray protons.

\section{Observations}\label{Sec:3}

In this Section, we introduce the observed regions, W44E and W44F, and present our observations of CO and $^{13}$CO species performed with the Atacama Pathfinder EXperiment (APEX\footnote{This publication is based on data acquired with the Atacama Pathfinder EXperiment (APEX). APEX is a collaboration between the Max-Planck-Institut f\"ur Radioastronomie, the European Southern Observatory, and the Onsala Space Observatory.} telescope, \citealt{Guesten06}). In Table~\ref{table1}, we provide observing parameters associated to each observed line (frequencies, beam sizes, corresponding sampling, and forward efficiencies).

\begin{table*}
\caption{Observed lines, and associated frequencies, beam sizes, sampling, and forward efficiencies.}             
\label{table1}      
\centering                          
\begin{tabular}{c  c  c  c  c  c   }        
\hline           
line & CO (3--2) & CO (4--3) & CO (6--5) & CO (7--6) & $^{13}$CO (3--2) \\
\hline
\hline
\footnotesize{$\nu$ (GHz)} & \footnotesize{345.796} & \footnotesize{461.041} & \footnotesize{691.473} & \footnotesize{806.652} & \footnotesize{330.588}\\
\footnotesize{FWHM ($''$)} & \footnotesize{18.2} & \footnotesize{13.5} & \footnotesize{9.0} & \footnotesize{7.7} & \footnotesize{18.9} \\
\footnotesize{sampling ($''$)} & \footnotesize{10} & \footnotesize{7} & \footnotesize{4} & \footnotesize{4} & \footnotesize{10} \\
\hline
\footnotesize{$F_{ \rm eff}$} & \footnotesize{0.97} & \footnotesize{0.95} & \footnotesize{0.95} & \footnotesize{0.95} & \footnotesize{0.97}\\
\hline
\end{tabular}
\end{table*}

\subsection{W44E}\label{Sub:3.1}
\subsubsection{Region}\label{Subsub:3.1.1}

As displayed in Fig. \ref{Fig:1}, W44E lies in the northeastern flattened part of the radio shell, where the SNR is interacting with the densest part of the molecular cloud (\citealt{Wootten:1977p16116}). Eastern of this region, \citet{Seta:2004p16955} located an "edge" in CO (1--0) emission, where the CO intensity drops by a factor of $\sim$2 and the emission lines are broadened and velocity-shifted with respect to the emission outside the SNR. They attribute this morphological feature to the interaction of the SNR with a clumpy molecular cloud, referring to the model of \citet{Chevalier:1999p16808}, where a radiative shell formed in the diffuse interclump gas drives slow molecular shocks into the clumps. The interaction between the giant molecular cloud and the SNR is believed to take place on the front side of the remnant as OH absorption is detected (\citealt{Reach:1998}). The drop in molecular column density is then attributed to dissociation of molecules and evaporation of molecular mass. \citet{Seta:2004p16955} did not detect wing emission towards W44E in CO (1--0), defined as lines with full velocity widths greater than 25 km s${}^{-1}$. Wing emission was however detected in CO (3--2) by \citet{Frail:1998p16306} and in CO (2--1) by {\citet{Reach:2000p16359} and \citet{Reach:2005p16099}. Both lines show a narrow component, tracing the unshocked gas also visible in ${}^{13}$CO (FWHM $\sim$5 km s${}^{-1}$, centered at $\sim$45 km s${}^{-1}$), and a broad component (FWHM $\sim$30 km s${}^{-1}$) tracing the shocked gas. Combining these observations with infrared spectroscopy, \citet{Reach:2000p16359} concluded to the existence of different shocks into moderate ($\sim$10$^2$ cm$^{-3}$) and high-density ($\sim$10$^4$ cm$^{-3}$) environments. In that picture, bright H${}_2$ emission together with emission of CO requires higher density gas residing in clumps that survived the initial blast wave. In a similar manner, \citet{Neufeld:2007p15887} identified five different groups of emission features towards W44E, classified on the basis of their spatial distribution. The pure rotational lines of H${}_2$ ($J>2$), together with lines of sulfur, constituted one of these groups, likely originating in molecular material subject to a slow, nondissociative shock.

In addition to the broad CO line emission, the observed H${}_2$ emission, and atomic fine structure lines, the existence of OH maser emission constitutes another strong indication of an interaction between the SNR and a molecular cloud. \citet{Claussen:1997p16194} detected ten OH maser features at 1720~MHz in W44E, two of them were shown to have multiple angular components (\citealt{Hoffman:2005p16236}). The masers appear along the continuum edges of the synchrotron emission. \citet{Claussen:1997p16194} interpret this region of enhanced synchrotron emission as featuring particle acceleration in a shock, referring to \citet{Blandford:1987p19995}. However, the enhancement could also be explained without the claim of new particle acceleration, originating from the compression of the magnetic field and pre-existing relativistic electrons (\citealt{Blandford1982,Frail:1998p16306}). The velocity of the masers have a low dispersion ($<$ 1~km~s${}^{-1}$) around a mean value of $\varv{}_{\rm LSR} = 44.7$~km~s${}^{-1}$, which agrees very well with the systemic velocity of W44E of $\varv{}_{\rm LSR} = 45$~km~s${}^{-1}$. The explanation for this low dispersion was given in terms of tangential amplification (\citealt{Frail:1996p19510}): the maser emission is only seen where the acceleration of the gas is transverse to the line of sight. From the strong inversion of the OH (1720~MHz) line through collisions with H${}_2$, possible excitation conditions can be inferred: kinetic temperatures should lie between 25 K $\le$ $T_{\rm k}$ $\le$ 200 K, while the density is expected to be 10${}^3$ cm${}^{-3}$ $\le$ $n_{\rm H_2}$ $\le$ 10${}^5$ cm${}^{-3}$ (\citealt{Elitzur:1976p19998}). \citet{Lockett:1999p19916} derived tighter physical conditions in the region, as they constrained a kinetic temperature of 50 K $\le$ $T_{\rm k}$ $\le$ 125 K, a molecular hydrogen density of 10${}^5$ cm${}^{-3}$, and OH column densities of $\sim$10${}^{16}$ cm${}^{-2}$. \citet{Frail:1998p16306} presented a map in CO (3--2) towards W44E, where they show that the masers delineate the forward edge of molecular gas, preferentially located nearer to the edge of the shock (as traced by the non-thermal emission) than the peak of CO. They report a correlation between the integrated CO (3--2) maps and the radio continuum.

   The line-of-sight magnetic field strength $B_{\parallel}$ in the region was determined by \citet{Claussen:1997p16194} to be within a factor 3 of 0.2 mG (with the expectation value of the total magnetic field given as $B = 2 B_{\parallel}$) in all maser locations using Zeeman splitting between the right- and left-circularly polarized maser lines in OH (1720 MHz). They found the direction of the field throughout the remnant to be constant.   \citet{Hoffman:2005p16236} constrained the strength of the magnetic field as  
   $\sim$1 mG on the basis of MERLIN and VLBA circular polarization observations of OH maser lines. The position angle of the magnetic field based on linear polarization of the masers aligns with the direction of the shocked dense gas filament.

\subsubsection{Data}\label{Subsub:3.1.2}
Observations towards the supernova remnant W44E were conducted in July 2010. We made use of a great part of the suite of heterodyne receivers available for this facility: FLASH345\footnote{This First Light APEX Submillimeter Heterodyne receiver was developed by Max Planck Institut f\"ur Radioastronomie, MPIfR, and commissioned in May 2010. In the 345 GHz band the dual-polarization receiver FLASH operates a 2SB SIS mixer provided by IRAM (Maier et al. 2005).}, FLASH460 (\citealt{Heyminck06}), and CHAMP$^+$ (\citealt{Kasemann06,Guesten08}), in combination with the MPIfR Fast Fourier Transform Spectrometer backend (FFTS, \citealt{Klein06}), or with the newly commissioned MPIfR X-Fast Fourier Transform Spectrometer backends (XFFTS, \citealt{Klein:2012}). The central position of all the observations was set to be $(\alpha_{[\rm{J}2000]}$ = 18$^{\rm h}$56$^{\rm m}$28 \fs 40, $\beta_{[\rm{J}2000]}$ = 01$^\circ$29$'$59 \farcs 0). Focus was checked at the beginning of each observing session, after sunrise and/or sunset on Mars, Jupiter, or Saturn. Continuum and CO line pointing was locally established on G34.26 and R-Aql. The pointing accuracy was found to be of the order of 5$''$ r.m.s, regardless of the receiver that was used. Table~\ref{table2} contains the main characteristics of the telescope, and of the observing set-up for each observed transition: used receiver, corresponding observing days, beam efficiency, system temperature, and spectral resolution.
The observations were performed in position-switching/raster mode using the APECS software (\citealt{Muders:2006}). The data were reduced with the CLASS software (see http://www.iram.fr/IRAMFR/GILDAS). 

\begin{table*}
\caption{Observed lines and corresponding telescope parameters -- the case of W44E.}             
\label{table2}      
\centering                          
\begin{tabular}{c  c  c  c  c  c   }        
\hline           
line & CO (3--2) & CO (4--3) & CO (6--5) & CO (7--6) & $^{13}$CO (3--2) \\
\hline
\hline
\footnotesize{receiver} & \footnotesize{FLASH345} & \footnotesize{FLASH460} & \footnotesize{CHAMP$^+$} & \footnotesize{CHAMP$^+$} & \footnotesize{FLASH345} \\
\footnotesize{observing days} & \footnotesize{13/07} & \footnotesize{13/07} & \footnotesize{14/07} & \footnotesize{14/07} & \footnotesize{15/07} \\
\hline
\footnotesize{$B_{ \rm eff}$} & \footnotesize{0.73} & \footnotesize{0.60} & \footnotesize{0.48} & \footnotesize{0.48} & \footnotesize{0.73} \\
\hline 
\footnotesize{$T_{\rm sys}$ (K)} & \footnotesize{162--205} & \footnotesize{436--573} & \footnotesize{1008--1643} & \footnotesize{2404--6455} & \footnotesize{192-228} \\
\footnotesize{$\Delta \varv$ (km s$^{-1}$)} & \footnotesize{0.066} & \footnotesize{0.476} & \footnotesize{0.635} & \footnotesize{0.544} & \footnotesize{0.069} \\    
\hline                                  
\end{tabular}\\
\end{table*}

\subsection{W44F}\label{Sub:3.2}
\subsubsection{Region}\label{Subsub:3.2.1}

The W44F region lies southeastern of W44E and hosts a thin filament of gas ranging from the northwest to the southeast, aligned with the radio continuum contours (see Fig. 1). According to the radio continuum emission, the long axis of the filament seems to be parallel to the shock front. \citet{Claussen:1997p16194} detected three OH 1720 MHz masers in this region with an average velocity of 46.6 km s${}^{-1}$. In the CO (3--2) map of \citet{Frail:1998p16306} the masers border the northern CO emission peak integrated over the red-shifted spectral wing between 49 km s${}^{-1}$ and 60 km s${}^{-1}$. The spectrum in CO (3--2) they observed towards W44F has a double-peak structure with peaks at 39 km s${}^{-1}$ and 50 km s${}^{-1}$. This double-peak structure is also visible in the spectrum of CO (2--1) (\citealt{Reach:2005p16099}). Although this could possibly be interpreted as a superposition of two moderately broad components in the line of sight, the deep trough's alignment with the peak in the ${}^{13}$CO (1--0) spectrum they also observed reveals that it actually is broad-line emission being absorbed by cold, narrow foreground gas. The magnetic field strenght was also measured in W44F using Zeeman splitting between the right- and left-circularly polarized maser lines in OH (1720 MHz) (\citealt{Frail:1998p16306,Hoffman:2005p16236}). The results were the same as in W44E, as well as the orientation of the position angle of the magnetic field being aligned with the filament of shocked gas.

\subsubsection{Data}\label{Subsub:3.2.2}

Observations towards the region W44F were conducted in several runs in the year 2009 (in June and August), and also in July 2012 (for ${}^{13}$CO (3--2)). The central position of all the observations was set to be $(\alpha_{[\rm{J}2000]}$ = 18$^{\rm h}$56$^{\rm m}$36 \fs 9, $\beta_{[\rm{J}2000]}$ = 01$^\circ$26$'$34 \farcs 6). Focus was checked at the beginning of each observing session, after sunrise and/or sunset on Mars, Jupiter, or Saturn. Continuum and line pointing was locally checked on G34.26 and R-Aql. The pointing accuracy was found to be of the order of 5$''$ r.m.s, regardless of the receiver that was used. Table~\ref{table3} contains the same parameters as Table~\ref{table2}, for observations of W44F. The observations were performed in position-switching/raster mode using the APECS software (\citealt{Muders:2006}). The data were reduced with the CLASS software (see http://www.iram.fr/IRAMFR/GILDAS). 

\begin{table*}
\caption{Observed lines and corresponding telescope parameters -- the case of W44F.}             
\label{table3}      
\centering                          
\begin{tabular}{c  c  c  c  c  c   }        
\hline           
line & CO (3--2) & CO (4--3) & CO (6--5) & CO (7--6) & $^{13}$CO (3--2) \\
\hline
\hline
\footnotesize{receiver} & \footnotesize{HET345} & \footnotesize{FLASH460} & \footnotesize{CHAMP$^+$} & \footnotesize{CHAMP$^+$} & \footnotesize{FLASH345} \\
\footnotesize{observing days} & \footnotesize{05/08} & \footnotesize{07/06} & \footnotesize{06/08} & \footnotesize{05/08} & \footnotesize{26/06/12} \\
\hline
\footnotesize{$B_{ \rm eff}$} & \footnotesize{0.73} & \footnotesize{0.60} & \footnotesize{0.52} & \footnotesize{0.49} & \footnotesize{0.73}  \\
\hline 
\footnotesize{$T_{\rm sys}$ (K)} & \footnotesize{278--284} & \footnotesize{410--484} & \footnotesize{1350--2024} & \footnotesize{3860--6079} & \footnotesize{256-337} \\
\footnotesize{$\Delta \varv$ (km s$^{-1}$)} & \footnotesize{0.106} & \footnotesize{0.318} & \footnotesize{0.635} & \footnotesize{2.178} & \footnotesize{0.035} \\    
\hline   
\end{tabular}\\
\end{table*}

\section{Averaged CO emission towards W44E and W44F}\label{Sec:4}

W44 is surrounded by several molecular clouds. To understand the CO emission towards W44E and W44F we have averaged all spectra obtained in each mapped region respectively, after convolving them to a common spatial resolution of $18 \farcs 2$, that of our CO (3--2) observations. These averaged spectra are shown in Fig. \ref{Fig:2}. The value for the systemic velocity of W44 that we found in the literature is 45 km s${}^{-1}$ (e.g. \citealt{Claussen:1997p16194}). The peaks of our spectra agree with this value. In the averaged spectra of W44F we detect strong absorption between $\sim$40 km~s${}^{-1}$ and $\sim$50 km~s${}^{-1}$ in all transitions up to CO (6--5), which is attributed to absorption by cold foreground gas.

\citet{Seta:1998p16736,Seta:2004p16955} have studied the vicinity of W44 in CO (1--0) and CO (2--1). In their earlier paper, presenting observations at low spatial resolution\footnote{Because \citet{Seta:1998p16736} were only able to separate structures larger than $\sim$15 pc, their observations cannot differentiate between W44E and F.}, they divided the CO emission towards W44 into three velocity components at $\varv{}_{\rm LSR} = 13$ km~s${}^{-1}$, 30--65 km~s${}^{-1}$, and 70--90 km~s${}^{-1}$. The 13 km~s${}^{-1}$ component corresponds to well-known foreground clouds in the solar neighborhood, in earlier publications attributed to the dust cloud Khavtassi3 at 15 km~s${}^{-1}$ (\citealt{Knapp:1974p16117,Wootten:1977p16116,Scoville:1987p17203}). We detect this component clearly in CO (3--2) towards both regions, although towards W44F it is mostly seen in absorption. In both fields, this component is rather spatially uniform, although towards W44E the intensity of this feature increases in the south-eastern direction. 

The 70--90 km s${}^{-1}$ component was attributed to clouds behind W44 with an uncertain fraction of emission from accelerated gas from W44\footnote{Wootten (1997) notes that there are no molecular absorption features in this velocity regime, therefore he concludes that the emitting gas is situated beyond the remnant.}. The integrated spectrum in CO (3--2) towards W44E shows a weak peak at 78 km s${}^{-1}$ that might corresponds to cloud 10 (CO G34.7--0.1, $\varv$=78 km s${}^{-1}$) in the labelling of \citet{Seta:1998p16736}, although the spatial distribution of the emission is difficult to disentangle from highly red-shifted shocked gas. Towards W44F we notice a broad emission feature in the velocity regime between 70 and 90 km s${}^{-1}$ in CO (3--2), peaking at 72 km s${}^{-1}$, together with a narrow peak at 86 km s${}^{-1}$ that is also seen in absorption in CO (4--3). Both features appear spatially uniform over the observed field.

The 30--65 km s${}^{-1}$ component is considered to be associated with W44. Towards W44E this central component appears even broader in our data (15-70 km s${}^{-1}$), and also towards W44F the wings seem to stretch further. \citet{Seta:1998p16736} identified six molecular clouds in this velocity range, of which three are candidates for an interaction (given their spatial coincidence with the SNR and similar estimated radial velocities) and one that is probably interacting with W44 (G34.8--0.6, $\varv$=48 km s${}^{-1}$), showing increased line widths and an abrupt shift in radial velocity at the rim of the SNR. At  54 km s${}^{-1}$ we detect another emission peak towards W44E, which could correspond to the absorption dip towards W44F at 54.5 km s${}^{-1}$. The emission at this velocity is strongly spatially varying. Its intensity increases towards the south of W44E, where it becomes comparable in intensity to the main peak. The origin might be a confined clump of material moving within the molecular cloud.

\section{CO maps}\label{Sec:5}   

\subsection{CO maps towards W44E}\label{Sub:5.1}

      \begin{figure*}
   \centering
     \includegraphics[width=\textwidth]{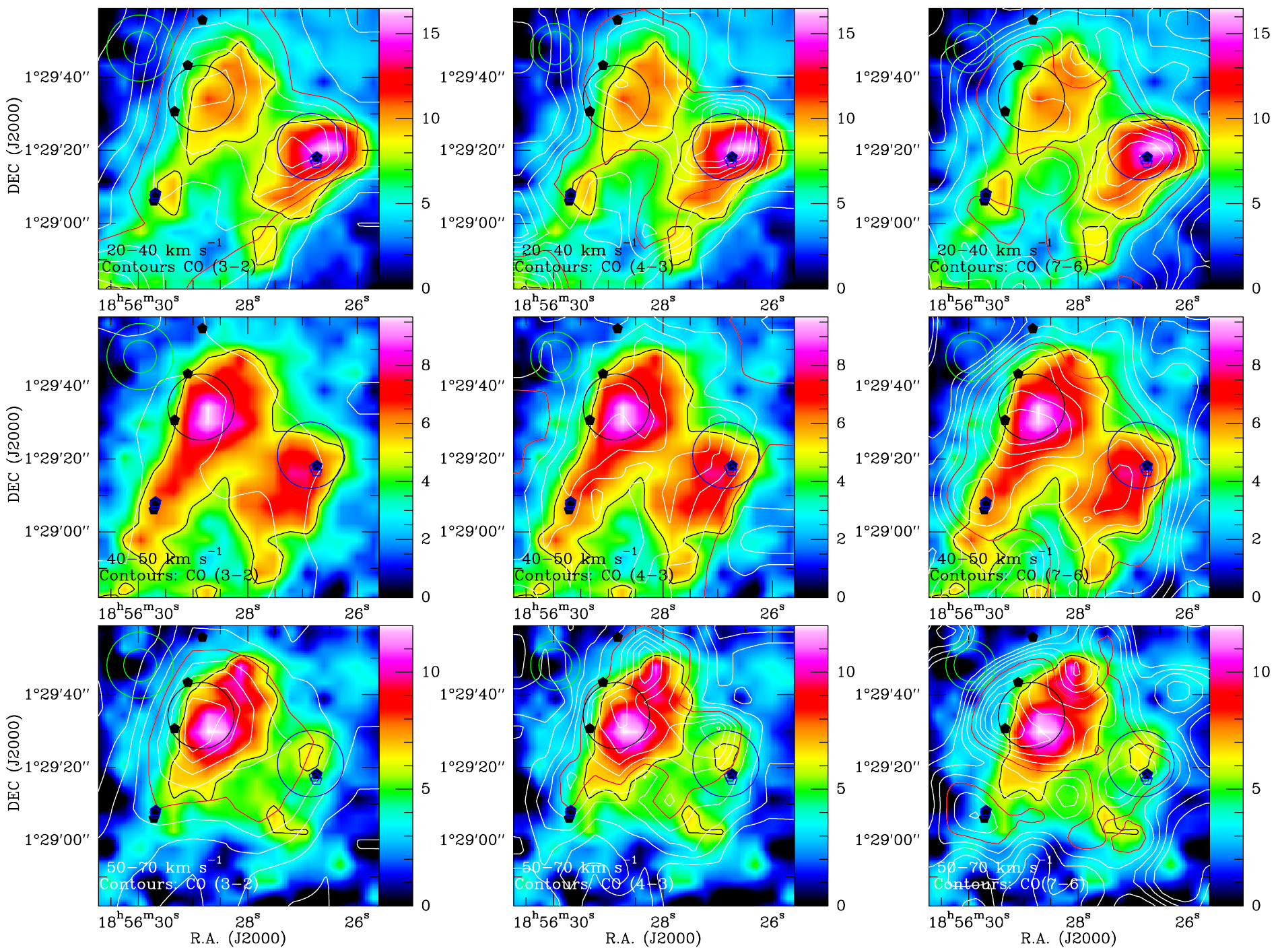}
      \caption{Overlays of the velocity-integrated maps of CO (3--2) (left column), CO (4--3) (middle column), and CO (7--6) (right column) as white contours on the CO (6--5) emission (colour background) observed towards W44E with the APEX telescope. The maps are in their original resolution (9 $\farcs 0$ for CO (6--5), 13 $\farcs 5$ for CO (4--3), and 18 $\farcs 2$ for CO (3--2)), except for CO (7--6), which was smoothed to the resolution of CO (4--3) in order to improve the signal-to-noise. The intensity was integrated between 20--40 km s${}^{-1}$ (blue wing, top row), 40--50 km s${}^{-1}$ (ambient emission, middle row), and 50--70 km s${}^{-1}$ (red wing, bottom row). The wedge unit is K km s${}^{-1}$ in antenna temperature. The contours are in steps of 10\%. The half-maximum contours of the colour and contour maps are indicated in red and black, respectively. The blue and black circles indicate the position of the our shock modelling analysis. the APEX beam sizes of the observations displayed are given in the upper left corner of each map. The black and blue hexagons mark the positions of the OH masers observed by \citet{Claussen:1997p16194} and \citet{Hoffman:2005p16236}. At the assumed distance of 3 kpc the 9 $\farcs 0$ beam of CO (6--5) corresponds to a spatial resolution of 0.13 pc.
                    }
         \label{Fig:3}
   \end{figure*}
   
        \begin{figure*}
   \centering
   \includegraphics[width=15 cm]{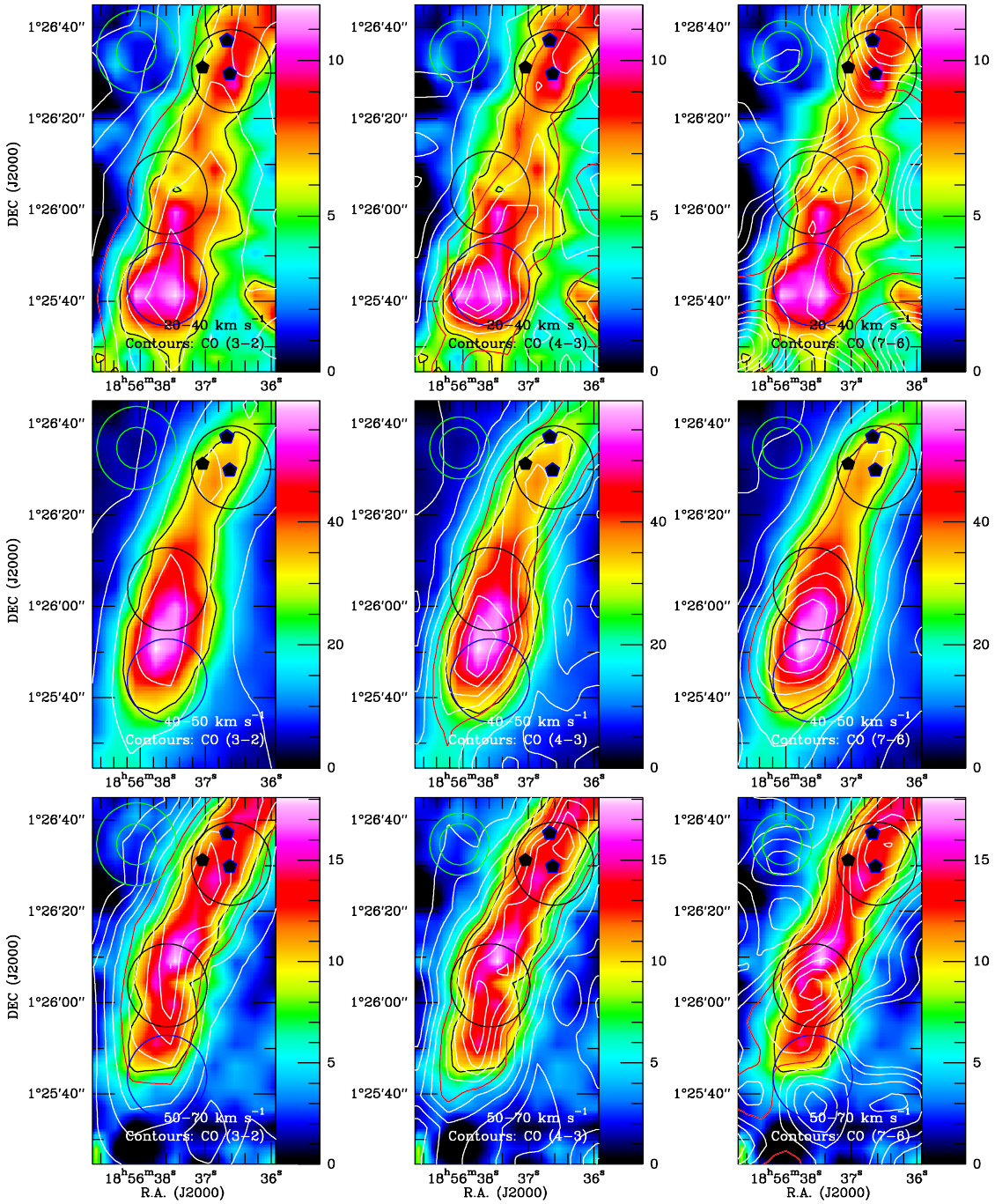}
      \caption{Same as Fig. 3 but observed towards W44F. The black and blue hexagons mark the positions of the OH masers observed by \citet{Claussen:1997p16194} and \citet{Hoffman:2005p16236}. 
                    }
         \label{Fig:4}
   \end{figure*}

 Figure \ref{Fig:3} shows maps of integrated CO emission\footnote{The velocity-integrated maps of CO shown in Fig. \ref{Fig:3} and \ref{Fig:4} are available as FITS files at the CDS via anonymous ftp to  xxx or via xxx .} in CO (6--5) with contours of integrated CO (3--2), (4--3), and (7--6) (respectively 1st, 2nd, and 3rd column) emission overlayed towards W44E, for the blue wing (top row, 20--40 km s${}^{-1}$), the red wing (bottom row, 50--70 km s${}^{-1}$), and the ambient velocity regime (central row, 40--50 km s${}^{-1}$). The positions of OH masers detected towards W44E are also displayed. The lower integration boundary of the blue wing of 20 km s${}^{-1}$ was chosen to avoid the contribution of the foreground cloud at $\sim$13 km s${}^{-1}$ in CO (3--2). Similarly we chose the upper integration boundary of the red wing as 70 km s${}^{-1}$ because at higher velocities the contribution of red-shifted gas related to W44 and background cloud emission reported by  \citet{Seta:1998p16736} is difficult to disentangle. 
 
 The emission distribution peaks in all CO maps behind or near the line of masers, which trace the shock front moving eastward into the ambient molecular gas (\citealt{Claussen:1997p16194}). The eastern peak exhibits strong wings towards higher and lower velocities relative to the systemic velocity of W44. There is another emission peak towards the interior of W44 close to the second masering region distinct from the delineated masers to the east. This peak is mostly seen in the blue-shifted velocity regime. Here the spatial displacement between maser and CO emission is less clear than for the other peak.

For our shock analysis, we aimed at choosing positions of clearly shocked gas, defined as positions of maximum CO emission in a given velocity range for a maximum number of observed transitions. Therefore we first determined the local maxima of CO emission in all of our transition maps in the various velocity regimes and then chose the positions of our analysis based on these maxima. In the centre panel of Fig. \ref{Fig:6} we show the two positions resulting from this procedure: the one close to the line of masers tracing the outer shock front (black circle, denoted as ``W44E1'') and the one towards the interior (blue circle, denoted as ``W44E2''). The coordinates of W44E1 and W44E2 are listed in Table \ref{table_pos}. 

\subsection{CO maps towards W44F}\label{Sub:5.2}

The corresponding maps of integrated CO emission for the three velocitiy regimes towards W44F, together with observed maser positions, are shown in Figure \ref{Fig:4}. The maps in the ambient velocity regime (central row) in CO (3--2) and CO (4--3) have to be treated with some caution as they suffer from substantial absorption in the line centres, as can e.g. be seen in Fig. \ref{Fig:2} and Fig. \ref{Fig:7}. In all velocity regimes, a thin filament of gas from the northwest to the southeast of the maps can be traced. Only the convolved map of CO (7--6) does not show such a clear filament in the blue velocity regime. The extent of the filament stretches further to the south for the blue wing emission than for the red wing. Again, we defined the position of clear shock interaction as showing maximum CO emission in a given velocity range for a maximum number of observed transitions. Accordingly, based on the local maxima of CO emission in all our transition maps, we selected three positions in W44F for our modelling analysis. These positions are displayed in the central panel of Fig. \ref{Fig:7}. The first position in the north covers the maser locations observed by \citet{Claussen:1997p16194} and \citet{Hoffman:2005p16236}. It is particularly prominent in the redshifted velocity regime (northern black circle in Fig. \ref{Fig:7}, denoted as ``W44F1''). There is a further maximum in the red wing regime (southern black circle in Fig. \ref{Fig:7}, denoted as ``W44F2''), while the emission integrated over the blue wing peaks still further to the south (blue circle in Fig. \ref{Fig:7}, denoted as ``W44F3''). The corresponding coordinates are listed in Table \ref{table_pos}.

\begin{table}
\caption{The positions of our analysis.}             
\label{table_pos}      
\centering                          
\begin{tabular}{l  l  l }    
           \noalign{\smallskip}    
\hline           
           \noalign{\smallskip}
Position &  $\alpha_{[\rm{J}2000]}$ & $\beta_{[\rm{J}2000]}$  \\
           \noalign{\smallskip}    
\hline
\hline
           \noalign{\smallskip}    

W44E1 & 18$^{\rm h}$56$^{\rm m}$28 \fs 90 & 01$^\circ$29$'$34 \farcs 2\\
W44E2  &18$^{\rm h}$56$^{\rm m}$26 \fs 90 & 01$^\circ$29$'$21\farcs 0\\
\hline
           \noalign{\smallskip}
W44F1  & 18$^{\rm h}$56$^{\rm m}$36 \fs 61 & 01$^\circ$26$'$30 \farcs 5\\
W44F2  & 18$^{\rm h}$56$^{\rm m}$37 \fs 58 & 01$^\circ$26$'$03 \farcs 8\\
W44F3  & 18$^{\rm h}$56$^{\rm m}$37 \fs 59 & 01$^\circ$25$'$43 \farcs 8\\
\hline
\end{tabular}\\
\end{table}

  \begin{figure*}
   \centering
   \includegraphics[width=\textwidth]{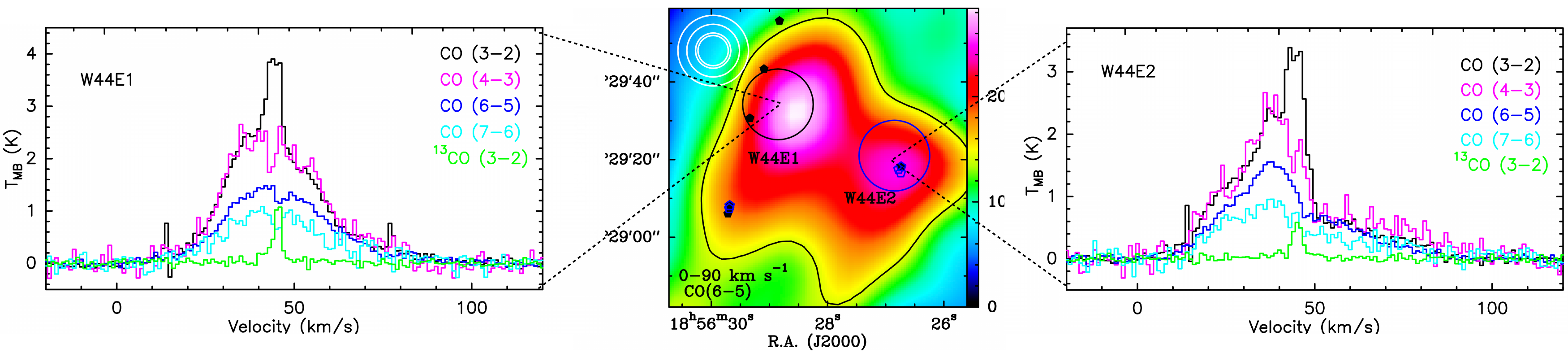}
      \caption{Central Panel: Positions of our analysis in W44E (W44E1: black circle, W44E2: blue circle) on the velocity-integrated CO (6--5) emission ($T_{\rm A}^*$ integrated between 0 and 90 km s${}^{-1}$) convolved to the CO (3--2) angular resolution of 18 $\farcs 2$ (colour background) with half-maximum contour in black. The wedge unit is K km s${}^{-1}$.  The blue and black circles indicate the APEX beam size of our observations in CO (3--2). The APEX beam sizes of our CO (3--2), CO (4--3), CO (6--5), and CO (7--6) observations are also provided (upper left corner, see also Table 1). The black and light blue hexagons mark the position of the OH masers observed by \citet{Claussen:1997p16194} and \citet{Hoffman:2005p16236}. Left Panel: Spectra observed in the position W44E1 (in $T_{\rm MB}$), CO (3--2), black; CO (4--3), pink; CO (6--5), dark blue; CO (7--6), light blue; ${}^{13}$CO (3--2), green. The CO spectral resolution is 1.0 km s${}^{-1}$ for all lines.  Right panel: Spectra observed in the position W44E2, same colours as in the left panel. }
         \label{Fig:6}
   \end{figure*}

   \begin{figure*}
   \centering
   \includegraphics[width=\textwidth]{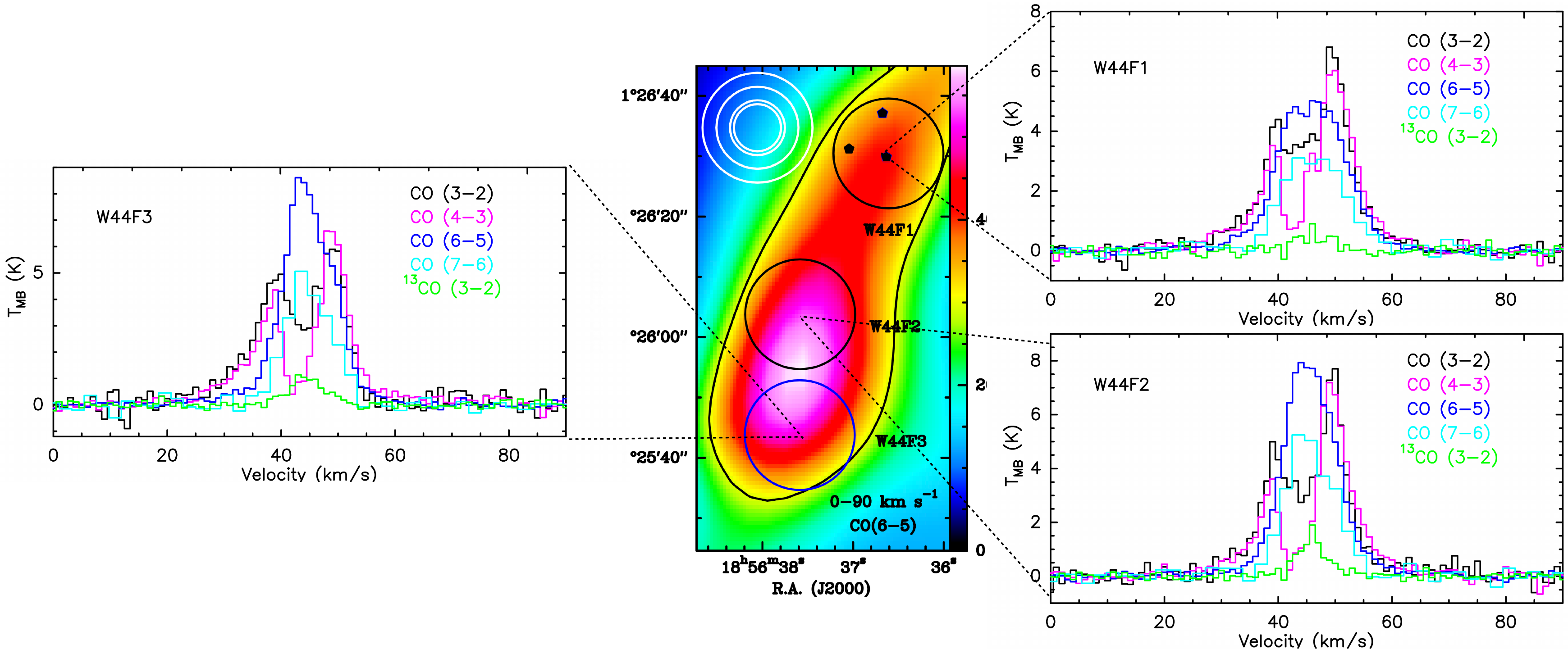}
      \caption{Same as Fig. \ref{Fig:6} but towards W44F. The positions in the central panel are W44F1, W44F2, and W44F3 (north to south). Left panel: Spectra observed in the position W44F3. Right panel: Spectra observed in the position W44F1 (top) and W44F2 (bottom). The CO spectral resolutions are 2.18 km s${}^{-1}$ for CO (7--6) and 1.0 km s${}^{-1}$ for all other lines.
              }
         \label{Fig:7}
   \end{figure*}

\section{Individual spectra of ${}^{12}$CO and ${}^{13}$CO}\label{Sec:6}

\subsection{Positions of analysis in W44E}\label{Sub:6.1}

The individual spectra towards our positions of analysis in W44E, as shown in Fig. \ref{Fig:6}, exhibit a complex structure. In CO (3--2), two components are seen as already observed by \citet{Frail:1998p16306}, \citet{Reach:2000p16359}, and \citet{Reach:2005p16099} in CO (3--2) and CO (2--1): there is a narrow component tracing the cold ambient gas and a broad component due to emission from shocked gas. 

Towards W44E1, the emission of the cold quiescent gas in ${}^{13}$CO (3--2) is confined between 40 and 50 km s${}^{-1}$ showing two components: one centred at 44.5 km s${}^{-1}$ with a FWHM of 5.2 km s${}^{-1}$ and another very narrow component (FWHM = 1 km s${}^{-1}$) at 45.6 km s${}^{-1}$. The FWHM of the former component is identical to the FWHM determined by \citet{Reach:2005p16099} for the central peak in CO (2--1) and its central velocity is close to the average OH (1720 MHz) maser velocity of 44.7 km s${}^{-1}$. The emission in ${}^{13}$CO aligns with the central, narrow peak in our spectrum of ${}^{12}$CO (3--2), although probable self-absorption renders the determination of the FWHM of the peak of the latter difficult. The lines in CO (4--3), (6--5), and (7--6) show self-absorption features in the range between 43 and 46 km s${}^{-1}$. The shock-broadened lines are asymmetric but similar in shape among all transitions. The red-shifted wings of the lines extent to high velocities at weak levels of emission, reaching well beyond the velocity regime of 30-65 km s${}^{-1}$ considered to be associated with W44 by \citet{Seta:1998p16736}.

Towards W44E2, the cold gas traced by ${}^{13}$CO (3--2) exhibits only one component centred at 44.8 km s${}^{-1}$ with an FWHM of 3.2 km s${}^{-1}$. This cold ambient component is also visible as a narrow peak in the spectrum of ${}^{12}$CO (3--2) at 44.6 km s${}^{-1}$. The shape of the shock-broadened line component is again similar in all transitions. Other than in W44E1, the main peak of the shock emission is slightly offset from the ambient emission at $\sim$38 km s${}^{-1}$. Furthermore two shoulders can be identified, one in the blue- and one in the red-shifted velocity regime, the latter again streching out to very high velocities of 80--90 km~s$^{-1}$.

Fig. \ref{Fig:5} shows the spectra of ${}^{12}$CO (3--2) and ${}^{13}$CO (3--2) towards W44E1 and W44E2 (two top panels, left ordinate), resampled to a common spectral resolution of 1 km~s$^{-1}$. The line temperature ratio of ${}^{12}$CO (3--2)/${}^{13}$CO (3--2) (right ordinate) is displayed in the lower velocity shift's regime, where ${}^{13}$CO is detected at more than 2$\sigma$ (orange dots) and 3$\sigma$ (red dots). Because these signal-to-noise values are only reached for $^{13}$CO towards the centres of the $^{12}$CO lines, the optical thickness values can accordingly only be inferred at the inner parts of the $^{12}$CO line wings. There, the line ratios yield optical thickness values of 3--7, assuming a typical interstellar abundance ratio of 50--60 (e.g., \citealt{Langer:1993p19291}). The line temperature ratio towards W44E at $\sim$13 kms$^{-1}$, corresponding to foreground emission, yields optical thickness values of 3--14 in this component. We note that this analysis relies on the assumption of an equal excitation temperature for both isotopologues.

\subsection{Positions of analysis in W44F}\label{Sub:6.2}

As already observed by \citet{Frail:1998p16306} and \citet{Reach:2005p16099}, the low velocity emission towards W44F seems to be absorbed by foreground gas in the lower transitions of CO. This effect is also present in our data, although additional absorption is caused by emission in our reference position. Compared to W44E, the shock emission appears narrower with FWHM less than 45 km s${}^{-1}$ and with the lines being more symmetric.

Towards W44F1, the quiescent gas as visible in ${}^{13}$CO (3--2) peaks at 46 km s${}^{-1}$ with an FWHM of 7.8 km s${}^{-1}$ (see Fig. \ref{Fig:5}), which agrees with the average OH (1720 MHz) maser velocity of 46.6 km s${}^{-1}$. Absorption is seen in all transitions to various extents in the lines of ${}^{12}$CO, the lowest transitions being affected in the full range between 39.5 and 50 km s${}^{-1}$. This is similar for W44F2, while there the emission of ${}^{13}$CO (3--2) shows a double-peaked structure with a broader component at 45.7 km s${}^{-1}$ (FWHM 6.7 km s${}^{-1}$) and a narrow component at 46.4 km s${}^{-1}$ (FWHM 0.7 km s${}^{-1}$). In W44F3, the narrow ambient component disappears again, the remaining component being centred at 44.4 km s${}^{-1}$ with an FWHM of 7.2 km s${}^{-1}$. W44F3 shows the most asymmetric profiles with a shoulder in the red-shifted wing (as visible in CO (6--5)) and pronounced emission in the blue wing. 

The line temperature ratio of ${}^{12}$CO (3--2)/${}^{13}$CO (3--2) towards W44F in the lower velocity regime, where ${}^{13}$CO is detected at more than 2$\sigma$ (orange dots) and 3$\sigma$ (red dots) is displayed in Figure \ref{Fig:5} (three bottom panels, right ordinate). As in W44E the ratios yield optical thickness values of 3--7 in the wings.

      \begin{figure}
   \centering
   \includegraphics[width=6.7cm]{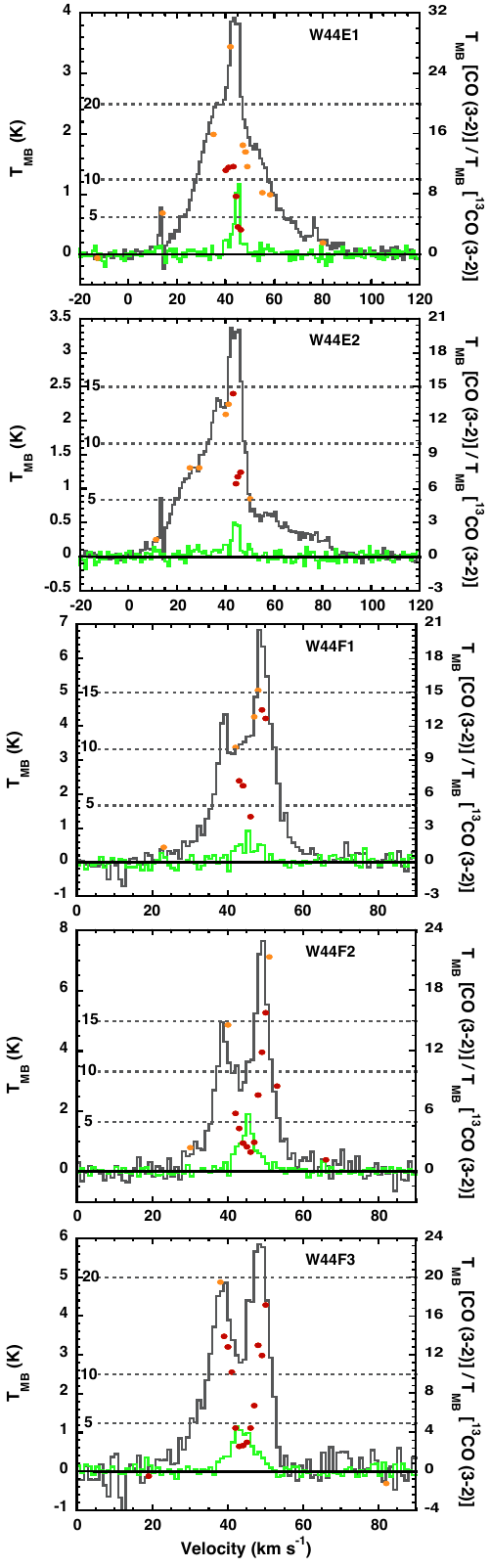}
      \caption{ Spectra (in $T_{\rm MB}$) of ${}^{12}$CO (3--2) (grey, left ordinate) and ${}^{13}$CO (3--2) (green, left ordinate) together with the line temperature ratio of  ${}^{12}$CO (3--2)/${}^{13}$CO (3--2) (dots, right ordinate). The ratio is shown for ${}^{13}$CO (3--2) line temperatures higher than 2$\sigma$ (orange dots) and 3$\sigma$ (red dots). Grey dotted lines indicate values of the temperature ratio as displayed on the left-hand side of these lines. The positions displayed are W44E1 and W44E2, and W44F1, W44F2, and W44F3 (top to bottom). The corresponding values of the optical thickness in the line wings are in the range of 3--7.
                     }
         \label{Fig:5}
   \end{figure}

\section{Modelling}\label{Sec:7}

\subsection{The observations}\label{Sub:7.1}
As described in Sect. \ref{Sec:6}, all spectra we have extracted towards our five positions of maximum CO emission in W44E and W44F show broad line wing emission around a central velocity consistent with the cold ambient gas. Most likely these profiles arise from a superposition of different shock components propagating into the dense gas of molecular clumps, which are nearly impossible to disentangle. In order to still get significant constraints on the dominant shock features and environmental conditions, we separated the profiles into a blue and a red lobe and applied our shock analysis to each of these velocity domains independently. 

All line profiles, convolved to a common angular resolution of 18$\farcs 2$, were fitted using a combination of up to four independent Gaussians with fully adjustable parameters. In W44E, these free fits yielded similar fit components for all transitions: two Gaussians in W44E1 and three in W44E2 with one additional ambient component in CO (3--2). In W44F, the fitting procedure was hindered by the deep absoption of the line centres. Therefore the fits for the CO (3--2) and CO (4--3) spectra could only be based on information from the wings, starting with initial values for the fitting procedure as obtained from the higher transitions. However, large uncertainties result from the lack of information for the fit of these lines. To handle this problem, we decided not to use the information of the line centres based on these fits, but rather to utilize the similar shape of emission lines among all transitions as observed in W44E. In particular, the ratio between the velocity-integrated intensities of the full lobe and the velocity-integrated intensities of only the wings turned out to be constant among all transitions within $\pm$5\% in W44E. Assuming a similar behaviour in W44F, we used this ratio derived from the CO (6--5) and CO (7--6) spectra to obtain estimations for the full-lobe integrated intensities in CO (3--2) and CO (4--3) based on their wing emission. 

The integration intervals were determined according to our knowledge of the ambient cloud component, based on the fits of ${}^{13}$CO (3--2) emission. To access the uncertainty of the lobe-separation with respect to the ambient velocity component, we varied the upper limit of the blue and the lower integration limit of the red lobe within the FWHM of the ${}^{13}$CO (3--2) fit. The integration intervals for our five positions are listed in Tables \ref{W44Eshock} and \ref{W44Fshock}. The errors for the integrated intensities in the CO (3--2) and CO (4--3) transitions in W44F, which were estimated based on the integrated intensities in the wings, were determined by varying the assumed lobe-wing integrated intensity ratio by $\pm$10 \%. For the final plot of the velocity-integrated intensities of all CO lines against the rotational quantum number of their upper level (so-called `spectral line energy distribution'), we also accounted for the filling factor of the lines, which is assumed to be the same for all transitions in each position, once the maps have all been convolved to the same resolution. We based these estimations on the half-maximum contours in the maps of CO (6--5), which provides high angular resolution, has a good signal-to-noise ratio, and which is barely affected by self-absorption. We note however that a reliable estimate of these factors would require high-resolution interferometric observations of the analyzed regions, which have not been performed so far. In ambiguous cases we examined various values of the filling factor within our analysis. The final values we obtained are listed in Tables \ref{W44Eshock} and \ref{W44Fshock}. To account for possible errors in the filling factor, we varied its value by $\pm$0.1.

The errors in the integrated line intensities due to uncertainties in the integration intervals and due to assumed filling factors were combined with the errors resulting from the inherent r.m.s noise fluctuations of the profiles and the uncertainties in subtracting baselines as described by \citet{Elfhag:1996p5257}. This results in the following total error in the integrated line intensities:
\begin{eqnarray}
\Delta I &=& \sqrt{\Delta I_{\rm line}^2+\Delta I_{\rm base}^2+\Delta I_{\rm integration}^2+\Delta I_{\rm FF}^2}\\ &=& \sqrt{\frac{\sigma}{\sqrt{N}}\cdot B \sqrt{\frac{\Delta \varv}{B-\Delta \varv}}+\Delta I_{\rm integration}^2+\Delta I_{\rm FF}^2}
\end{eqnarray}
where $\Delta \varv$ (km s${}^{-1}$) is the total velocity width of the spectral line, $N$ is the total number of spectrometer channels covering a total velocity range of $B$ km s${}^{-1}$, and $\sigma$ is the r.m.s noise per channel. 

As can be seen in the rotational diagrams displayed in Fig. \ref{W44Efits} and \ref{W44Ffits}, the large errors in the column densities of the upper levels of the (4--3) and (3--2) transitions in W44F mirror the uncertainties in the flux reconstruction necessary in order to account for the absorption in the line centre. The errors in the spectral line energy distributions and the rotational diagrams, respectively, are larger in W44F than in W44E because in W44E the ratio between flux in the line wings and flux in the line centre is higher, such that the uncertainty due to the unknown value of the ambient velocity is of less consequence.

\subsection{The models}\label{Sub:7.2}

For the interpretation of the observed spectral energy distributions (`SLEDs'), we used a combined model consisting of a one-dimensional, plane-parallel shock model and a radiative transfer module (\citealt{Gusdorf:2008p974,Gusdorf:2012p18545}). The former model solves the magneto-hydrodynamical equations in parallel with a large chemical network (including more than 100 species linked by over 1000 reactions) for stationary C- and J-type shocks (\citealt{Flower:2003p1558,Flower:2003p1795}) or approximated non-stationary CJ-type shocks. The theoretical elements for the computing of approximated, non-stationary shock models have been introduced in \citet{Chieze:1998} and \citet{Lesaffre:2004a,Lesaffre:2004b}. They have been proven useful mostly in the interpretation of H$_2$ pure rotational lines observed in shocks associated to bipolar outflows that accompany star formation, (e.g. \citealt{Giannini:2004,Gusdorf:2008p974,Gusdorf:2011}) and associated to supernova remnants (e.g. \citealt{Cesarsky:1999}).

The radiative transfer of H${}_{2}$ as the main gas coolant is self-consistently calculated for the first 150 ro-vibrational levels within all shock models. Input parameters are the pre-shock density $n_{\rm H}$, the shock velocity $\varv_{\rm s}$, and the magnetic field parameter value $b$, given as $B$[$\mu$G] = $b \times \sqrt{n_{\rm H}[{\rm cm}^{-3}]}$, as well as the shock age for non-stationary shocks. The dynamical variables and chemical fractional abundances calculated with this model for each point through the shock wave are then passed on to a radiative transfer module, which calculates the emission from the CO molecule based on the 'large velocity gradient' (LVG) approximation. The integrated line temperatures of the first 40 rotational transitions of CO can then be compared to the observed SLED, as described in \citet{Gusdorf:2012p18545}.

{\footnotesize
\begin{table*}[h]
\caption{ Shock model parameters. Grid intervals $\Delta x$ are given as minimum and maximum increments found in the grid.}             
\label{Shockparam}     
\centering                          
\begin{tabular}{l  c  c  c  c  c  c  c  c }        
\hline           
Shock type & number of models & $\varv$ [km s$^{-1}$] & $\Delta \varv$ [km s${}^{-1}$] & $b$ & $\Delta b$&  $n_{\rm H}$ [cm${}^{-3}$] & age [yr] & $\Delta$age  [yr]\\
\hline
\hline
            \noalign{\smallskip}
 C-type\tablefootmark{a} & 98 & 20 -- 55 &  2 -- 5&  0.45 -- 2 &0.15 -- 0.25 &10$^3$, 10$^4$, 10$^5$, 10$^6$ &   -- &  --\\
 C-type\tablefootmark{b}  &  9 &  20 -- 40 &  10 &  1 -- 3 &  0.5 &  10$^5$ & -- & --\\
CJ-type &  1004 & 10 -- 50 & 2 --5 & 0.3 --2 &  0.15 -- 0.25 &  10$^4$, 10$^5$ &  4 -- 8895 &  1 -- 1285\\
\hline
            \noalign{\smallskip}

\end{tabular}
\tablefoottext{a}\citet{Gusdorf:2008p974}
\tablefoottext{b}\citet{Anderl:2013}

\end{table*}}

Our grid of models consists of more than 1000 integrated intensity diagrams obtained for C- and CJ-type shock models\footnote{The grid also contains high-density models subject to grain-grain processing as calculated by \citet{Anderl:2013}.}.  
The range of model parameters covered by our grid is listed in Table \ref{Shockparam}. We note that the coverage of our grid is not completely uniform over the range parameters, therefore the table gives maximum und minimum grid intervals for each parameter as found in our grid. The observed SLEDs were compared with this grid using a $\chi^2$ routine to select the best fits. In the fits, we only used the most unambiguous shock-tracing lines CO (7--6) and CO (6--5), which are not expected to be optically thick in the line wings and which are not affected by self-absorption. In W44E, a further reason for that choice is the existence of background emission unrelated to W44 in the low-lying CO transitions, which is expected in the velocity range 70-90 km s${}^{-1}$ (see Sect. \ref{Sec:4}). Because the redshifted line wings stretch much further in W44E than in W44F, it is impossible to discern this background emission from shocked gas related to W44E. In W44F this problem does not exist due to the narrower line widths, but here the integrated intensities in CO (3--2) and CO (4--3) are only rough estimates based on the line wings and can therefore only serve as additional constraint for the fits based on the higher transitions.

  \begin{figure}
   \centering
   \includegraphics[width=8cm]{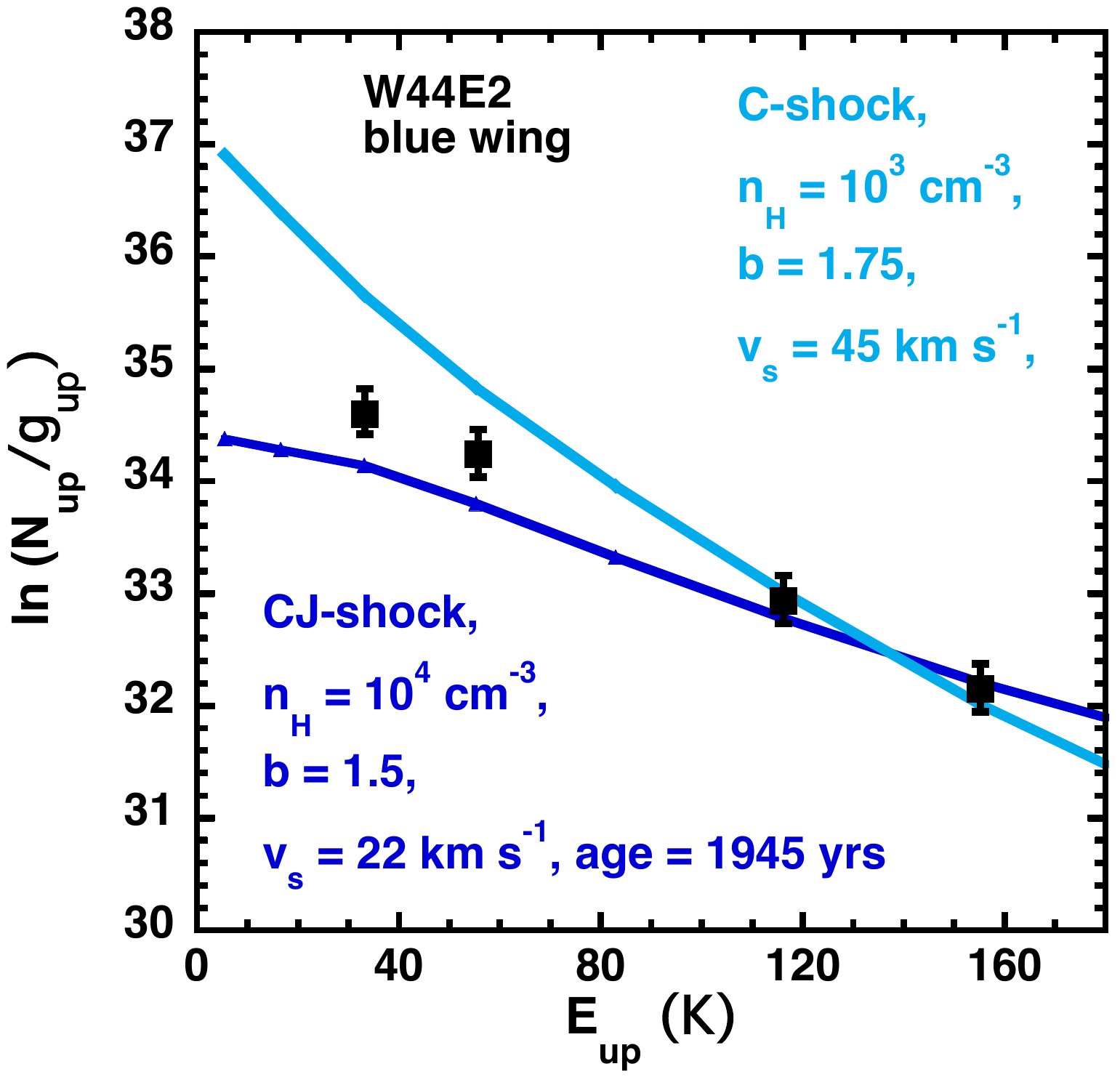}
      \caption{ Rotational diagram of the blue shock component in W44E2. Observations are marked by black squares. The lines show the C-type (dark blue) and CJ-type (light blue) shocks that yield the best fits to the CO (7--6) and (6--5) transitions.
          }
         \label{Fig:C-CJ}
   \end{figure}

   \begin{figure*}[!Htb]
   \centering
   \includegraphics[width=\textwidth]{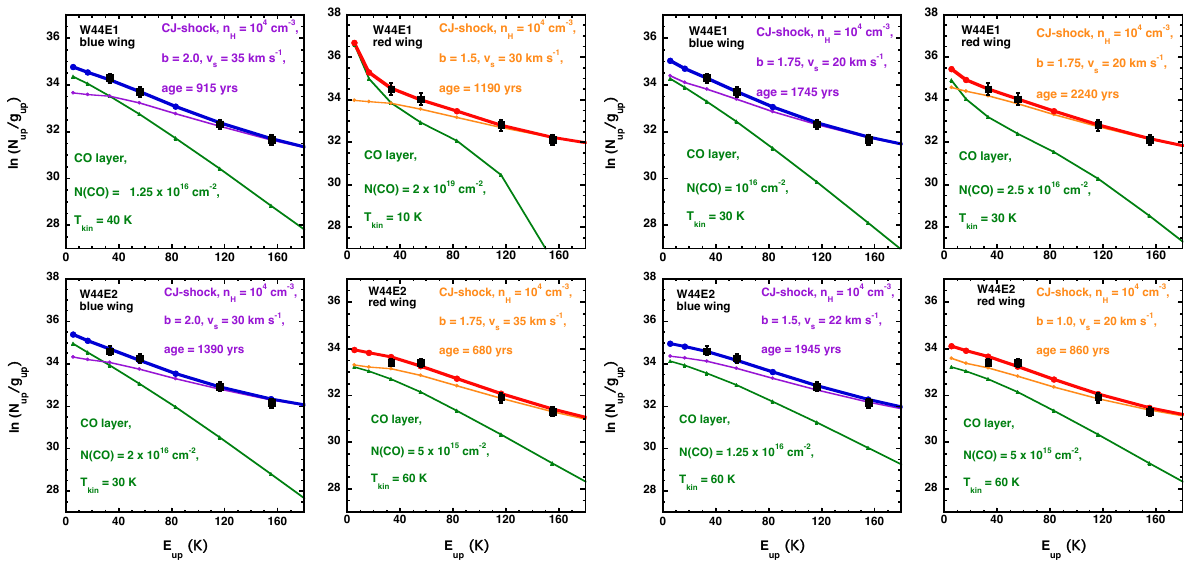}
      \caption{ Best-model comparisons between CO observations and models for all positions in W44E (W44E1: top row, W44E2: bottom row) for both the blue and red shock components, respectively. The four panels on the left show the high velocity fits while the four panels on the right show the low velocity fits in each position. Observations are marked by black squares, the best-shock models are displayed in violet line (blue lobe) and orange line (red lobe) and diamonds, the CO layer that we used to compensate the ambient emission affecting the CO (3--2) and CO (4--3) transitions in green line and triangels, and the sum of ambient and shock component in blue line (blue lobe) and red line (red lobe) with circles.               }
         \label{W44Efits}
   \end{figure*}
   
   \begin{figure}[!Htb]
   \centering
   \includegraphics[width=9cm]{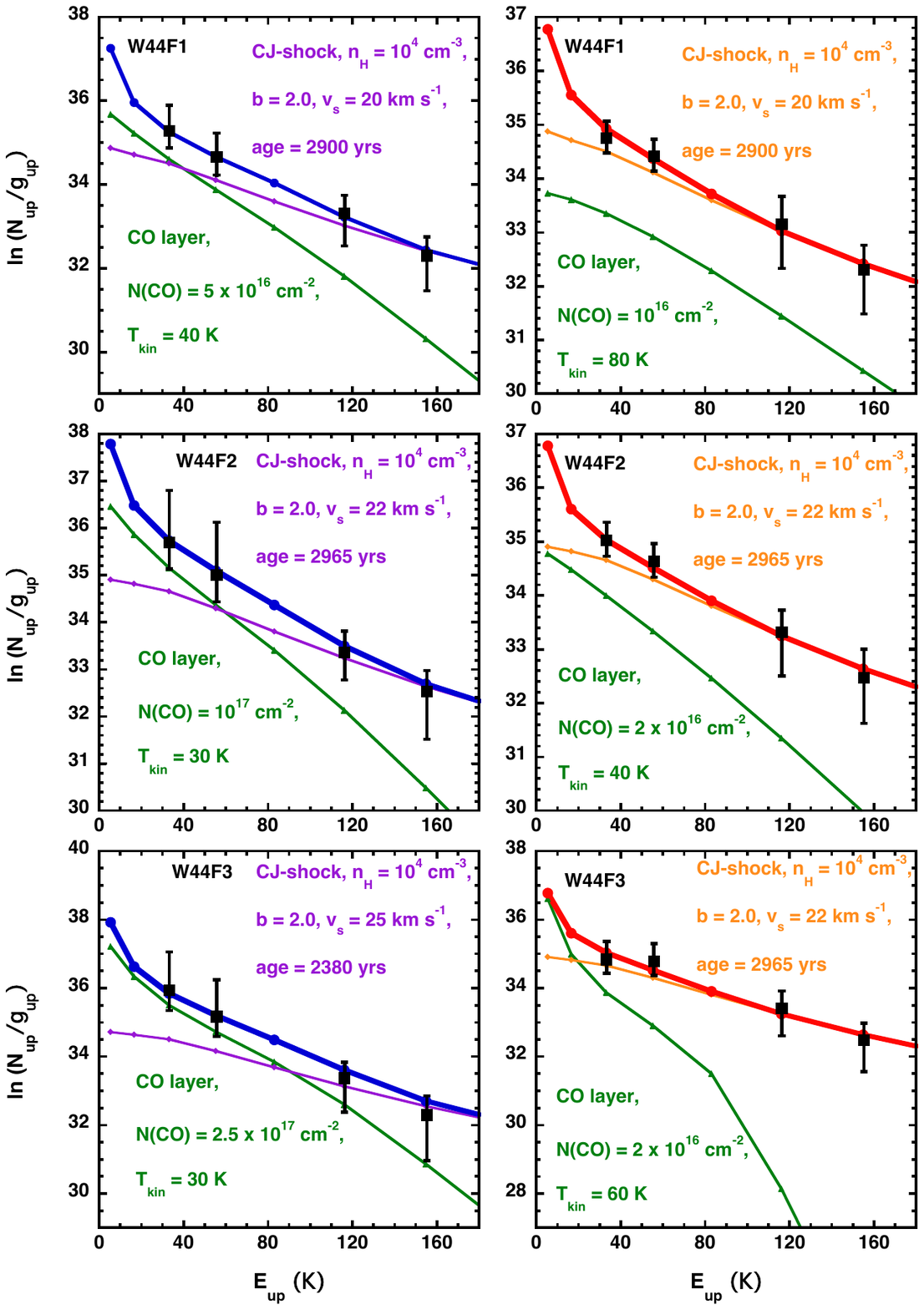}
      \caption{ Best-model comparisons between CO observations and models for all positions in W44F (W44F1: top row, W44F2: centre row, W44F3: bottom row) for both the blue (left) and red (right) shock components, respectively. The colour code is the same as in Fig. \ref{W44Efits}.               }
         \label{W44Ffits}
   \end{figure}

   \begin{figure}
   \centering
   \includegraphics[width=9cm]{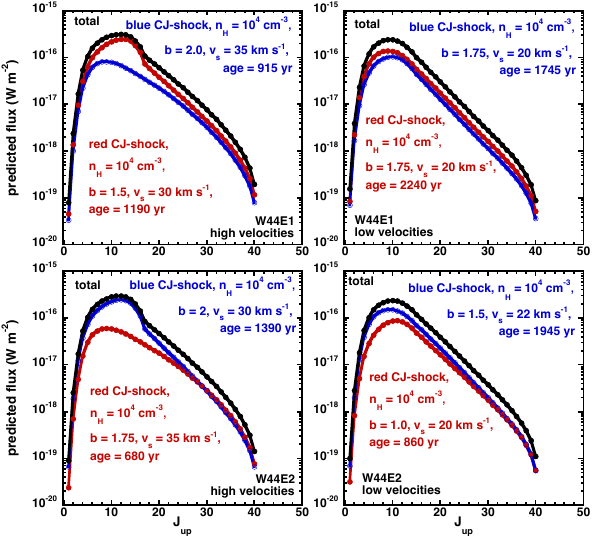}
      \caption{ Integrated CO fluxes (W m$^{-2}$) against the rotational quantum numbers of the upper level in W44E. The highest level of $J_{\rm up}=40$ corresponds to a maximum energy of 4513 K. Displayed are the integrated fluxes for W44E1 as predicted by the high-velocity models (left panel) and low-velocity models (right panel) for W44E1 (top row) and W44E2 (bottom row).
      }
         \label{Fig:10}
   \end{figure}

   \begin{figure}
   \centering
   \includegraphics[width=5cm]{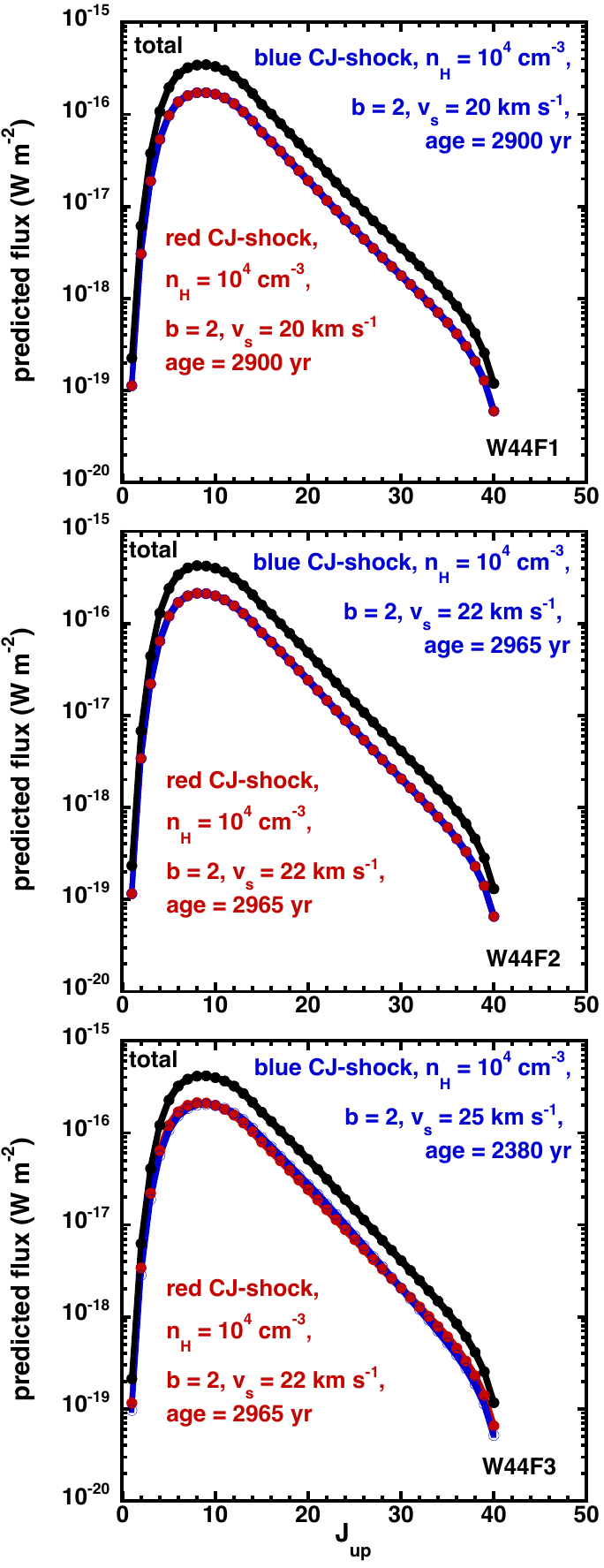}
      \caption{ Integrated CO fluxes (W m$^{-2}$) against the rotational quantum numbers of the upper level in W44F. The highest level of $J_{\rm up}=40$ corresponds to a maximum energy of 4513 K. Displayed are the integrated fluxes for the positions W44F1 (top), W44F2 (middle), and W44F3 (bottom).
          }
         \label{Fig:11}
   \end{figure}

\section{Results}\label{Sec:8}

\subsection{Shock emission}

\subsubsection{W44E}

Our $\chi ^2$ comparison of the observed integrated intensities of CO in the unambiguously shock tracing transitions (7--6) and (6--5) towards W44E with our grid of models showed the observations to be compatible with non-stationary CJ-shocks. 
Such shocks have not yet reached a state of dynamical equilibrium. \citet{Lesaffre:2004a,Lesaffre:2004b} have proven that such young shocks can be approximated by introducing a J-type discontinuity in a C-type flow at a point in the steady-state profile that is located increasingly downstream as the age of the shock advances, hence their designation as \lq CJ-type'. Given their younger shock age, the calculation is subsequently stopped sooner in the post-shock gas. The result is that the overall profile is found to be warmer, but also thinner than that of a stationary C-type shock, resulting in less CO emission, with a characteristic shift of the excitation towards higher-lying transitions. For the blue lobe in W44E2, the integrated line intensities in CO (7--6) and CO (6--5) were found to be consistent with either CJ-type models with a pre-shock density of $n_{\rm H} = 10^4$~cm$^{-3}$, or with C-type ones, with a pre-shock density of $10^3$~cm$^{-3}$, as can be seen in Fig.~8. Because of the wider and colder thermal profile generated in C-type shocks, it can be seen that the modelled integrated intensities in CO(3--2) and CO(4--3) of a C-type shock strongly overtop the observed ones. For this reason, and because the 10$^3$~cm$^{-3}$ pre-shock density significantly differs from the derived densities in the other positions, we discarded this fitting solution. 
J-type shocks, on the other hand, do not match our observations because they would predict a higher relative excitation of the higher CO transitions due to the higher temperatures produces in these discontinuous shocks. Based on these considerations, we conclude that young CJ-type shocks seem to be most consistent with our observations. We note that these models are also successful in predicting the optical thickness (thinness) of the lower (upper) lying CO transitions.

Given the presumably young age of the shocks in the molecular clumps close to the edge of the remnant (being much younger than the assumed SNR's age of 20,000 years), this result seems reasonable. The pre-shock density is found to be 10${}^{4}$ cm${}^{-3}$, consistent with previous estimates for the dense clump component (\citealt{Reach:2000p16359,Reach:2005p16099}). The total, blue- to redshifted velocity extent of the CO lines towards W44E, which is $\sim$60 km s${}^{-1}$, imposes a constraint on possible shock velocities: the gas must be accelerated to velocities $\ge$ 30 km s${}^{-1}$. 
 Obeying this constraint we obtain projected shock velocities of 35 and 30 km s${}^{-1}$ for the blue and the red lobe in W44E1, respectively, while in W44E2 we get 30 and 35 km s${}^{-1}$ for the blue and the red lobe. The pre-shock magnetic field component perpendicular to the shock layers in our models is constrained to $B =$ 200 $\mu$G for the blue lobes and 150 or 175 $\mu$G for the red lobes in W44E1 and W44E2. The difference could either be attributed to uncertainties in the analysis, be interpreted as a projection effect, or a as a slight inhomogeneity of the medium. The age of the shocks scatters around 1000 years, where the shocks of the dominant lobes for each position (red in W44E1 and blue in W44E2) appear older than the shocks in the reverse directions. In general the whole double-peak structure in W44E bears some bipolar characteristics, where antiparallel shocks exhibit similar parameters, pointing towards a general uniformity of the medium over the observed scales along the line of sight.

We note that the best fits to the observed SLEDs, except for the red lobe in W44E2, were achieved with slow CJ-shocks at $\sim$20 km s${}^{-1}$, which are much less consistent with the broad linewidths. We however also list them (together with the best fit in the 20 km s${}^{-1}$ velocity range towards W44E2, red lobe) in Table \ref{W44Eshock}, as one could argue that they might constitute the dominant shock component as compared to possible faster but weaker components. They agree with the faster solutions in terms of the pre-shock density and the magnetic field strength value, but show higher ages of almost 2000 years.

The post-shock density of all models fitting the observations in W44E is of the order of $9 \times 10^{4}$ -- $1.5 \times 10^5$ cm${}^{-3}$, yielding expected post-shock values of the magnetic field for this component of 390--720 $\mu$G, compatible with the magnetic field strength estimated from maser circular polarization observations (see Sect. \ref{Subsub:3.1.1}). The models produce OH column densities of a few $10^{14}$ cm${}^{-2}$ while the CO column density in the models is of the order of $4 \times 10^{16}$ -- $2 \times 10^{17}$ cm${}^{-2}$.

\subsubsection{W44F}

The line widths in W44F are narrower than in W44E. Accordingly, we find that the observations are compatible with CJ-type shocks with velocities between 20 and 25 km s${}^{-1}$ as best fits to our observations (cf. Table \ref{W44Fshock}). The pre-shock magnetic field component perpendicular to the shock layers in our models is given by $b =$2 corresponding to 200 $\mu$G and the pre-shock density given as 10${}^{4}$ cm${}^{-3}$, consistent with W44E. The age of the shocks in W44F is $\sim$2900 years. This result is somewhat puzzling given the much younger age obtained for W44E. The difference is less strong if we consider the slow shock solutions in W44E. However, the different line-widths in W44E and F already hint at a different nature of the shocks in both regions, probably linked to inhomogeneities in the interstellar medium. The post-shock density of the models consistent with the CO emission in W44F is of the order of $7 \times 10^{4}$ -- $9 \times 10^4$ cm${}^{-3}$, similar to W44E, with corresponding magnetic field values of $\sim$600 $\mu$G. The OH column densities are of the order of $7 \times 10^{14}$ -- $1 \times 10^{15}$ cm${}^{-2}$, while for CO the column densities are $\sim 3 \times 10^{17}$ cm${}^{-2}$.

\subsection{Degeneracies of the shock models}

To get an idea of the robustness and degeneracy of the results presented in the previous section, we have examined the best-fitting models, having $\chi^2$ values of at most twice the smallest value, for each position in both velocity lobes. Except for the two C-type models in W44E2 already mentioned above and some C-type cases in W44F with relatively high values of $\chi^2$, all models are CJ-type. Furthermore all these CJ-models agree in a pre-shock value of 10${}^{4}$ cm${}^{-3}$. 

In W44E, the mean values of the pre-shock magnetic field strengths of the best-fitting models lie within the range of $\sim$150--175 $\mu$G with standard deviations of about 40~$\mu$G\footnote{A somewhat special case is W44E2, red lobe, where the best fitting, high-velocity model and the second best model have $\chi^2$ values that differ by a factor of 2.3, such that there is no considerable fitting degeneracy.}. The mean velocity values in W44E are located between 22 (W44E1, red lobe) and 35 km s$^{-1}$ (W44E2, red lobe) with a maximum standard deviation of 9 km s$^{-1}$ (W44E1, blue lobe). The largest scatter among positions and velocity lobes in W44E is found with respect to the ages of the shocks, where the mean values are 1700 and 680 years in W44E2, blue and red lobe respectively, and 1350 and 2000 years in W44E1, blue and red lobe respectively, with standard deviations of $\sim$400 years. However, as will be discussed later, the fitting of the red lobe emission in W44E is subject to increased uncertainties.

In W44F, the best-fitting models lie closer together in terms of their values of $\chi^2$ than in W44E. The range of mean values of the pre-shock magnetic field strengths of the best-fitting models in W44F is the same as in W44E with standard deviations of 30--50 $\mu$G. The mean velocity values lie between 24 and 27 km s$^{-1}$ with standard deviations of $\sim$5 km s$^{-1}$. The mean ages of the best-fitting models lie in the range of 2050--2750 years, with standard deviations of the order of 700 years.

\subsection{Unshocked CO layers}

   \begin{figure}
   \centering
   \includegraphics[width=\columnwidth]{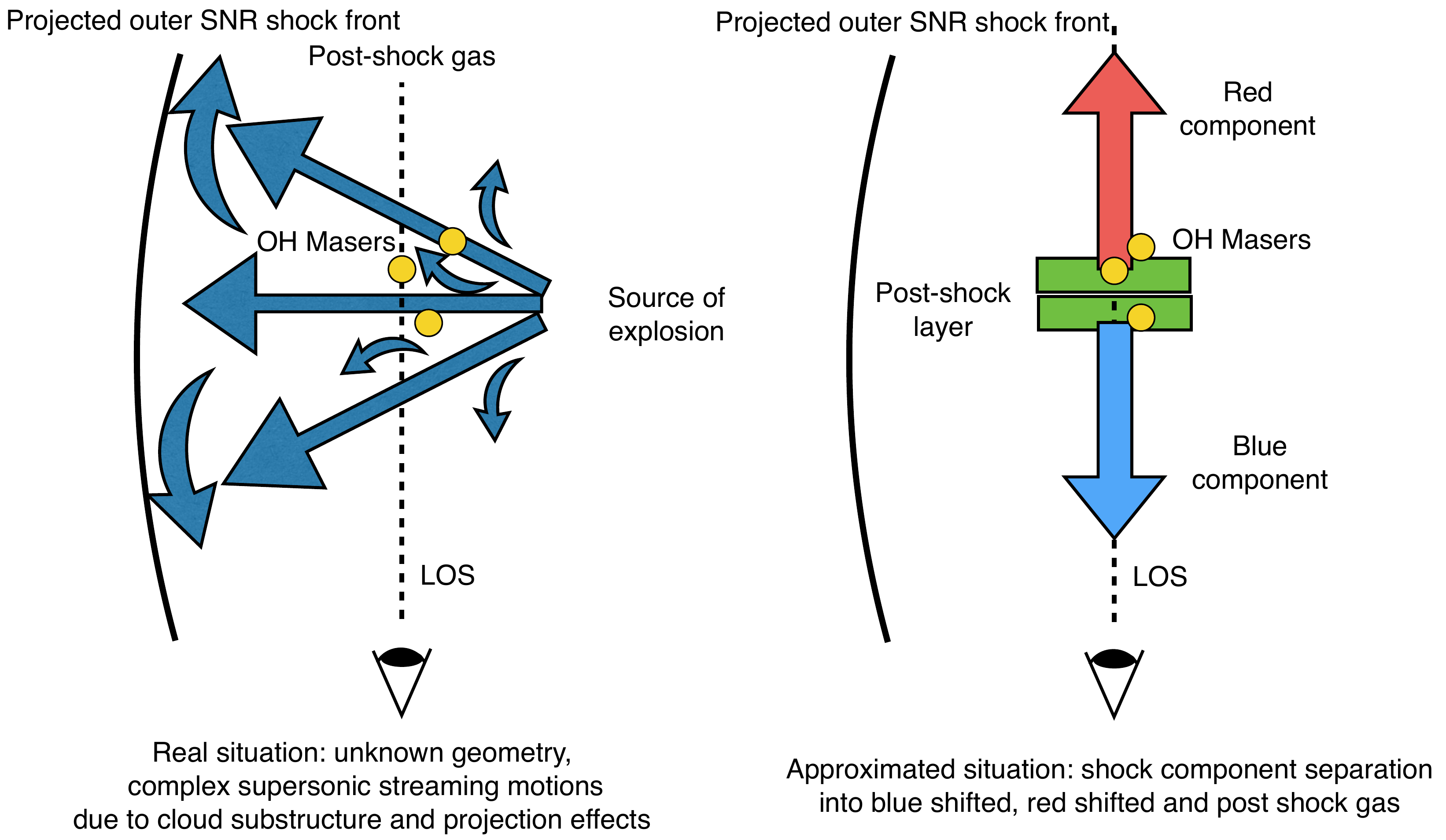}
      \caption{  Schematic illustration of our modelling approach. The main SNR shock direction is perpendicular to the line of sight as traced by OH masers, which are known to reside in dense postshock gas. The spectral line broadening, however, hints at supersonic fluid motion along the line of sight, which are expected, given projection effects and complex cloud substructure (left sketch). Our modelling decomposes the complex shock geometry in three components: one shock layer propagating away from the observer (fitting the red-shifted component of the line profiles), one shock layer propagating towards the observer (fitting the blue-shifted emission of the line profiles), and an additional layer of gas (found to correspond to postshock conditions) that extends into the line wings (right sketch). This latter layer makes up for the unrealistic accumulation of matter in the post-shock region inherent to 1D shock modelling, which does not allow the post-shock gas to relax in all possible directions, as well as for existing post-shock gas created and stirred up by shocks perpendicular to the line-of-sight. 
          }
         \label{Fig:comic}
   \end{figure}

As already found for W28F (\citealt{Gusdorf:2012p18545}), all fitting shock models underestimate the emission in CO (3--2) and CO (4--3). This is not too surprising as we do not expect our shock models to accurately model the dense and cold postshock gas due to their one-dimensional geometry. Furthermore, the existence of OH masers hints at the presence of dense postshock gas that is due to a shock transverse to the line-of-sight and thus different from the line-wing shock emission we consider in our shock models (See Fig. \ref{Fig:comic}). We therefore added a layer of CO to the shock models, calculated with the LVG module in `homogeneous slab' mode. This was done by calculating a large grid with varying density ($n_{\rm H}=$ 10${}^4$, 5$\times$10${}^4$, and 10${}^5$ cm${}^{-3}$), temperature (10--150 K), line width ($\Delta \varv$=1, 5, and 10 km s$^{-1}$), CO fractional abundance (10${}^{-4}$--10${}^{-6}$), and CO column density (2$\times$10$^{13}$--1$\times$10$^{20}$ cm${}^{-2}$). This grid of 2700 models was then compared to the residuals of the shock-modelled integrated intensities in CO (3--2) and CO (4--3) using a $\chi^2$ routine. The results are listed in Tables \ref{W44Elayer} and \ref{W44Flayer}. 

In W44E, the layers complementing the shock fits are at low and moderate temperatures between 10 and 60 K with densities always higher than the obtained preshock density of 10${}^4$ cm${}^{-3}$. The CO column densities vary between 5$\times$10$^{15}$ and 2$\times$10$^{19}$ cm${}^{-2}$. 
The best-model comparisons of the CO observations and the shock and layer components are displayed in Fig. \ref{W44Efits}. The sum of both components gives an encouraging fit to the data except for the integrated intensity of CO (3--2) and (4--3) in the red lobe of the W44E2 position. As we have noted before, the corresponding spectrum extends extremely far in the red-shifted range of the spectrum, showing only weak emission and is therefore prone to possible uncertainties in the baseline subtraction. Given the expected contamination by background emission, a poor baseline quality in CO (4--3), and the fact that our grid of shock models is not well suited for these high shock velocities, it might not be too surprising that our fit was not as successful in this case. 

The CO layers complementing the shock models in W44F show temperature values between 30 and 80 K. The densities, as in W44E, are higher than the obtained preshock density in all added layers. The layer temperatures in the blue wings are always lower than in the red wings. The inferred CO column densities have values between $10^{16}$ and 2.5$\times$$10^{17}$ cm${}^{-2}$. The quality of the fits, displayed in Fig. \ref{W44Ffits}, is satisfying in all integrated intensity diagrams. Given their density and temperature values, the added gas layers in W44E and W44F are dominated by postshock gas. We however note that, given the geometric uncertainties, a more detailed interpretation of these layers (e.g. possible contamination by unshocked ambient / foreground gas) has to remain open.

\section{Discussion}\label{Sec:9}

SNRs consitute an important source of energy and momentum input for the interstellar medium. Tables \ref{W44Einjection} and \ref{W44Finjection} show the mass, momentum, and energy injection calculated per beam for our best-fitting shock models. The shocked mass per beam amounts to $\sim$1  and $\sim$2 $M_{\sun}$ for the models in W44E and F, respectively. The total energy input in one beam, given as the sum of the blue- and the red-lobe shock contributions, amounts to $\sim$1$\times$10$^{46}$ erg in W44E and $\sim$1.5$\times$10$^{46}$ erg in W44F. This contribution approximately corresponds to the fraction of the total initial energy of 10$^{51}$ erg that is expected to be ejected into the observed area by the supernova, assuming an isotropic explosion. This would mean that the interaction of the SNR with molecular clouds constitutes regions of high energy dissipation. The values of the momentum input as sum of the red- and blue-shifted velocity lobes in one beam lies between $\sim$4$\times$10$^{6}$ $M_{\sun}$ cm s$^{-1}$ and $\sim$7$\times$10$^{6}$ $M_{\sun}$ cm s$^{-1}$, while in W44F it is $\sim$1$\times$10$^{7}$ $M_{\sun}$ cm s$^{-1}$. The results for all positions in both velocity intervals are listed in Tables \ref{W44Einjection} and \ref{W44Finjection}.

Figures \ref{Fig:10} and \ref{Fig:11} show the integrated CO fluxes (in W m$^{-2}$) within our 18$\farcs$2 beam for all rotational transitions up to an upper rotational quantum number of 40 for all best-fitting shock models (cf. \citealt{Hailey:2012}). Displayed are the shock models that fit the red- and blue-wing emission (red and blue dots, respectively) together with the total line integrated fluxes as sum of both wings (black dots). The asymmetry of the line profiles in W44E is mirrored in the plots, as well as the extensive symmetry of the lines in W44F. As can be seen from Figure  \ref{Fig:10}, observations of higher-lying CO transitions will provide an opportunity to further remove modelling degeneracies, in particular the degeneracy between slow and fast models in W44E, because the location of the peak in the CO-ladder differs for different models. Such observations will be possible with the SOFIA telescope. Given that our fits are based on only two transitions that unambiguously trace shock emission, it is very important to add more lines to our analysis.

Finally, we must note that the extraction of the integrated intensities relies on some assumptions with respect to the shock geometry in the analyzed region. Following the method introduced in the study of \citet{Gusdorf:2012p18545}, we split the whole spectral line into blue- and red-shifted parts without completely omitting the central line emission. 
The real situation, however, is most certainly much more complex. The existence of maser spots hints at the fact that a dominant shock direction is perpendicular to the line-of-sight, with the corresponding post-shock emission thus also being found in the line centres. The broad line wings might then stem from the projected wings of a bow-shock, with the ambient medium being compressed and pushed aside (for a schematic illustration see Fig. \ref{Fig:comic}).

Our analysis can only be applied to shock components propagating parallel to the line-of-sight and therefore gives averaged information on the shock conditions only in the gas being pushed towards or away from us. The derived properties, as long as they are non-directional such as the total magnetic field strength or the density, should still be valid, though. However, it is important to keep in mind that the geometrical assumptions we applied are severe and a more accurate modelling in terms of geometry would need to account for the influence of the shock component perpendicular to the line-of-sight as well. At the same time, we note that the high level of microphysical detailedness included in our shock model cannot be maintained within fully two- or three-dimensional models. An alternative would be the use of pseudo-multidimensional models, synthesized from one-dimensional shocks, as presented by e.g. \citet{Kristensen:2008} and \citet{Gustafsson:2010}, but in any case, a multi-dimensional modelling approach would add a set of additional, barely constrained parameters, such as the orientation of the magnetic field with respect to the shock structure, and therefore increase the degeneracy of the modelling in a prohibitive way.

\begin{table*}
\caption{Modelling results for W44E: shock emission.}             
\label{W44Eshock}     
\centering                          
\begin{tabular}{c  c  c  c  c  c  c  c  c  c}        
\hline           
Position & integration interval & filling factor & shock type & $b$ & velocity [km s${}^{-1}$] & $n_{\rm H}$ [cm${}^{-3}$] & age [yr]\\
\hline
\hline
\footnotesize{ W44E1} & \footnotesize{(-20 to 44.5 km s${}^{-1}$)} & \footnotesize{0.75} & \footnotesize{CJ} & \footnotesize{2} & \footnotesize{35} & \footnotesize{10${}^{4}$}  & \footnotesize{915} \\
 & \footnotesize{\it low velocity fit:}& \footnotesize{\it } & \footnotesize{\it CJ} & \footnotesize{\it 1.75} & \footnotesize{\it 20} & \footnotesize{\it 10${}^{4}$}  & \footnotesize{\it 1745} \\
 & \footnotesize{(44.5 to 120 km s${}^{-1}$)} & \footnotesize{0.5} & \footnotesize{CJ} & \footnotesize{1.5} & \footnotesize{30} & \footnotesize{10${}^{4}$} &  \footnotesize{1190}\\
& \footnotesize{\it low velocity fit:}& \footnotesize{\it } & \footnotesize{\it CJ} & \footnotesize{\it 1.75} & \footnotesize{\it 20} & \footnotesize{\it 10${}^{4}$}&  \footnotesize{\it 2240} \\
\hline
\footnotesize{ W44E2} & \footnotesize{(-20 to 44.8 km s${}^{-1}$)} & \footnotesize{0.5} & \footnotesize{CJ} & \footnotesize{2} & \footnotesize{30} & \footnotesize{10${}^{4}$} &  \footnotesize{1390} \\
& \footnotesize{\it low velocity fit:}& \footnotesize{\it } & \footnotesize{\it CJ} & \footnotesize{\it 1.5} & \footnotesize{\it 22} & \footnotesize{\it 10${}^{4}$}&  \footnotesize{\it 1945} \\
 & \footnotesize{(44.8 to 120 km s${}^{-1}$)} & \footnotesize{0.75} & \footnotesize{CJ} & \footnotesize{1.75} & \footnotesize{35} & \footnotesize{10${}^{4}$}&  \footnotesize{680} \\
 & \footnotesize{\it low velocity fit:}& \footnotesize{\it } & \footnotesize{\it CJ} & \footnotesize{\it 1.0} & \footnotesize{\it 20} & \footnotesize{\it 10${}^{4}$} &   \footnotesize{\it 860} \\
\hline
\end{tabular}\\
\end{table*}

\begin{table*}
\caption{Modelling results for W44E: CO layers.}             
\label{W44Elayer}     
\centering                          
\begin{tabular}{c  c  c  c  c  c  c}        
\hline           
Position & integration interval & $n_{\rm H}$ [cm${}^{-3}$] & $\Delta \varv$ [km s${}^{-1}$]& $T_{\rm kin}$ (K)& N(CO) [cm${}^{-2}$] & X(CO)\\
\hline
\hline
\footnotesize{ W44E1} & \footnotesize{(-20 to 44.5 km s${}^{-1}$)} & \footnotesize{$5\cdot10^4$} & \footnotesize{5} & \footnotesize{40} & \footnotesize{$1.25\cdot 10^{16}$} & \footnotesize{$5\cdot10^{-6}$} \\
 & \footnotesize{\it low velocity fit:}& \footnotesize{\it $1\cdot10^5$} & \footnotesize{\it 10} & \footnotesize{\it 30} & \footnotesize{\it 2 $\cdot$ 10${}^{16}$}& \footnotesize{\it 1 $\cdot$ 10$^{-5}$}   \\
 & \footnotesize{(44.5 to 120 km s${}^{-1}$)} & \footnotesize{$1\cdot10^5$} & \footnotesize{10} & \footnotesize{10} & \footnotesize{$2\cdot 10^{19}$}& $5\cdot10^{-5}$  \\
& \footnotesize{\it low velocity fit:}& \footnotesize{\it $5\cdot10^4$} & \footnotesize{\it 1} & \footnotesize{\it 30} & \footnotesize{\it 2.5 $\cdot$ 10${}^{16}$}& \footnotesize{\it 5 $\cdot$ 10$^{-6}$}   \\
\hline
\footnotesize{ W44E2} & \footnotesize{(-20 to 44.8 km s${}^{-1}$)} & \footnotesize{$1\cdot10^5$} & \footnotesize{10} & \footnotesize{30} & \footnotesize{$2.5\cdot 10^{16}$} & $1\cdot10^{-5}$ \\
& \footnotesize{\it low velocity fit:}& \footnotesize{\it $5\cdot10^4$} & \footnotesize{\it 5} & \footnotesize{\it 60} & \footnotesize{\it 1.25 $\cdot$ 10${}^{16}$}  & \footnotesize{\it 5 $\cdot$ 10$^{-6}$} \\
 & \footnotesize{(44.8 to 120 km s${}^{-1}$)} & \footnotesize{$5\cdot10^4$} & \footnotesize{10} & \footnotesize{60} & \footnotesize{$5\cdot 10^{15}$}& $5\cdot10^{-6}$  \\
 & \footnotesize{\it low velocity fit:}& \footnotesize{\it $5\cdot10^4$} & \footnotesize{\it 10} & \footnotesize{\it 60} & \footnotesize{\it 5 $\cdot$ 10${}^{15}$}& \footnotesize{\it 5 $\cdot$ 10$^{-6}$}  \\
\hline
\end{tabular}\\
\end{table*}

\begin{table*}
\caption{Modelling results for W44F: shock emission.}             
\label{W44Fshock}     
\centering                          
\begin{tabular}{c  c  c  c  c  c  c  c  c c}        
\hline           
Position & integration interval & filling factor & shock type & $b$ & velocity [km s${}^{-1}$] & $n_{\rm H}$ [cm${}^{-3}$] &  age [yr]\\
\hline
\hline
\footnotesize{ W44F1} & \footnotesize{(0 to 46.2 km s${}^{-1}$)} & \footnotesize{0.5} & \footnotesize{CJ} & \footnotesize{2} & \footnotesize{20} & \footnotesize{10${}^{4}$}&  \footnotesize{2900} \\
 & \footnotesize{(46.2 to 90 km s${}^{-1}$)} & \footnotesize{0.5} & \footnotesize{CJ} & \footnotesize{2} & \footnotesize{20} & \footnotesize{10${}^{4}$}&  \footnotesize{2900} \\
\hline
\footnotesize{ W44F2} & \footnotesize{(0 to 45.7 km s${}^{-1}$)} & \footnotesize{0.5} & \footnotesize{CJ} & \footnotesize{2} & \footnotesize{22} & \footnotesize{10${}^{4}$}  & \footnotesize{2965} \\
 & \footnotesize{(45.7 to 90 km s${}^{-1}$)} & \footnotesize{0.5} & \footnotesize{CJ} & \footnotesize{2} & \footnotesize{22} & \footnotesize{10${}^{4}$}&   \footnotesize{2965} \\
\hline
\footnotesize{ W44F3} & \footnotesize{(0 to 44.4 km s${}^{-1}$)} & \footnotesize{0.5} & \footnotesize{CJ} & \footnotesize{2} & \footnotesize{25} & \footnotesize{10${}^{4}$}  & \footnotesize{2380} \\
 & \footnotesize{(44.4 to 90 km s${}^{-1}$)} & \footnotesize{0.5} & \footnotesize{CJ} & \footnotesize{2} & \footnotesize{22} & \footnotesize{10${}^{4}$} &  \footnotesize{2965} \\
\hline
\end{tabular}\\
\end{table*}

\begin{table*}
\caption{Modelling results for W44F: CO layers.}             
\label{W44Flayer}     
\centering                          
\begin{tabular}{c  c  c  c  c  c  c}        
\hline           
Position & integration interval & $n_{\rm H}$ [cm${}^{-3}$] & $\Delta \varv$ [km s${}^{-1}$]& $T_{\rm kin}$ (K)& N(CO) [cm${}^{-2}$]& X(CO)\\\hline
\hline
\footnotesize{ W44F1} & \footnotesize{(0 to 46.2 km s${}^{-1}$)} & \footnotesize{$5\cdot10^4$} & \footnotesize{5} & \footnotesize{40} & \footnotesize{$5\cdot 10^{16}$} & \footnotesize{$1\cdot10^{-5}$} \\
 & \footnotesize{(46.2 to 90 km s${}^{-1}$)} & \footnotesize{$5\cdot10^4$} & \footnotesize{10} & \footnotesize{80} & \footnotesize{$1\cdot 10^{16}$} & \footnotesize{$1\cdot10^{-5}$} \\
\hline
\footnotesize{ W44F2} & \footnotesize{(0 to 45.7 km s${}^{-1}$)} & \footnotesize{$1\cdot10^5$} & \footnotesize{10} & \footnotesize{30} & \footnotesize{$1\cdot 10^{17}$} & \footnotesize{$1\cdot10^{-5}$}\\
 & \footnotesize{(45.7 to 90 km s${}^{-1}$)} & \footnotesize{$1\cdot10^5$} & \footnotesize{10} & \footnotesize{40} & \footnotesize{$2\cdot 10^{16}$} & \footnotesize{$1\cdot10^{-5}$} \\
\hline
\footnotesize{ W44F3} & \footnotesize{(0 to 44.4 km s${}^{-1}$)} & \footnotesize{$5\cdot10^4$} & \footnotesize{10} & \footnotesize{30} & \footnotesize{$2.5\cdot 10^{17}$} & \footnotesize{$5\cdot10^{-5}$}\\
 & \footnotesize{(44.4 to 90 km s${}^{-1}$)} & \footnotesize{$1\cdot10^5$} & \footnotesize{10} & \footnotesize{60} & \footnotesize{$2\cdot 10^{16}$} & \footnotesize{$1\cdot10^{-5}$}\\
\hline
\end{tabular}\\
\end{table*}

\begin{table*}
\caption{Mass, momentum and energy injection per beam for W44E.}             
\label{W44Einjection}     
\centering                          
\begin{tabular}{c  c  c  c  c c }        
\hline           
Position & integration interval & filling factor & mass [$M_{\sun}$]  & momentum [$M_{\sun}\, $cm s$^{-1}$] & energy [erg]\\ % 
\hline
\hline
\footnotesize{ W44E1} & \footnotesize{(-20 to 44.5 km s${}^{-1}$)} & \footnotesize{0.75} & \footnotesize{1.04} & \footnotesize{3.78$\cdot$10${}^{6}$} & \footnotesize{7.59$\cdot 10^{45}$} \\
 & \footnotesize{\it low velocity fit:}& \footnotesize{\it 0.75} & \footnotesize{\it 1.21} & \footnotesize{\it 2.55 $\cdot$ 10${}^6$} & \footnotesize{\it 3.54 $\cdot$ 10${}^{45}$} \\
 \footnotesize{ } & \footnotesize{(44.5 to 120 km s${}^{-1}$)} & \footnotesize{0.5} & \footnotesize{1.01} & \footnotesize{3.26$\cdot$10${}^{6}$} & \footnotesize{4.59$\cdot 10^{45}$} \\
 & \footnotesize{\it low velocity fit:}& \footnotesize{\it 0.5} & \footnotesize{\it 1.54} & \footnotesize{\it 3.30 $\cdot$ 10${}^6$} & \footnotesize{\it 4.34 $\cdot$ 10${}^{45}$} \\
\hline
\footnotesize{ W44E2} & \footnotesize{(-20 to 44.8 km s${}^{-1}$)} & \footnotesize{0.5} & \footnotesize{1.41} & \footnotesize{4.42$\cdot$10${}^{6}$} & \footnotesize{7.64$\cdot 10^{45}$} \\
 & \footnotesize{\it low velocity fit:}& \footnotesize{\it 0.5} & \footnotesize{\it 1.25} & \footnotesize{\it 2.99 $\cdot$ 10${}^6$} & \footnotesize{\it 3.73 $\cdot$ 10${}^{45}$} \\
 \footnotesize{ } & \footnotesize{(44.8 to 120 km s${}^{-1}$)} & \footnotesize{0.75} & \footnotesize{0.76} & \footnotesize{2.73$\cdot$10${}^{6}$} & \footnotesize{5.25$\cdot 10^{45}$} \\
 & \footnotesize{\it low velocity fit:}& \footnotesize{\it 0.75} & \footnotesize{\it 0.57} & \footnotesize{\it 1.21 $\cdot$ 10${}^6$} & \footnotesize{\it 1.18 $\cdot$ 10${}^{45}$} \\
\hline
\end{tabular}\\
\end{table*}

\begin{table*}
\caption{Mass, momentum and energy injection per beam for W44F.}             
\label{W44Finjection}     
\centering                          
\begin{tabular}{c  c  c  c  c c }        
\hline           
Position & integration interval & filling factor & mass [$M_{\sun}$]  & momentum [$M_{\sun}\, $cm s$^{-1}$] & energy [erg]\\ % 
\hline
\hline
\footnotesize{ W44F1} & \footnotesize{(0 to 46.2 km s${}^{-1}$)} & \footnotesize{0.5} & \footnotesize{1.96} & \footnotesize{4.27$\cdot$10${}^{6}$} & \footnotesize{6.01$\cdot 10^{45}$} \\
 \footnotesize{ } & \footnotesize{(46.2 to 90 km s${}^{-1}$)} & \footnotesize{0.5} & \footnotesize{1.96} & \footnotesize{4.27$\cdot$10${}^{6}$} & \footnotesize{6.01$\cdot 10^{45}$} \\
\hline
\footnotesize{ W44F3} & \footnotesize{(0 to 45.7 km s${}^{-1}$)} & \footnotesize{0.5} & \footnotesize{2.23} & \footnotesize{5.28$\cdot$10${}^{6}$} & \footnotesize{7.54$\cdot 10^{45}$} \\
 \footnotesize{ } & \footnotesize{(45.7 to 90 km s${}^{-1}$)} & \footnotesize{0.5} & \footnotesize{2.23} & \footnotesize{5.28$\cdot$10${}^{6}$} & \footnotesize{7.54$\cdot 10^{45}$}\\
 
 \hline
\footnotesize{ W44F3} & \footnotesize{(0 to 44.4 km s${}^{-1}$)} & \footnotesize{0.5} & \footnotesize{1.89} & \footnotesize{5.08$\cdot$10${}^{6}$} & \footnotesize{7.83$\cdot 10^{45}$} \\
 \footnotesize{ } & \footnotesize{(44.4 to 90 km s${}^{-1}$)} & \footnotesize{0.5} & \footnotesize{2.23} & \footnotesize{5.28$\cdot$10${}^{6}$} & \footnotesize{7.54$\cdot 10^{45}$}\\

\hline
\end{tabular}\\
\end{table*}

\section{Conclusions}\label{Sec:10}

\begin{enumerate}
\item We have presented new mapping observations with the APEX telescope in $^{12}$CO (3--2), (4--3), (6--5), (7--6) and $^{13}$CO (3--2) towards regions E and F in W44. The spectra, averaged over the respective regions, exhibit emission features at 13 km s$^{-1}$, corresponding to foreground emission, and emission between 70 and 90 km s$^{-1}$, which is at least partly due to background emission (\citealt{Seta:1998p16736}). The averaged lines show broad wing emission, between $\sim$15 and $\sim$90 km~s$^{-1}$ in W44E and between $\sim$25 and $\sim$60 km s$^{-1}$ in W44F. Accordingly the red-shifted emission in W44E is hard to distentangle from background emission.
\item The velocity integrated maps in the blue (20--40 km s$^{-1}$) and red (50--70 km s$^{-1}$) velocity regimes in W44E show one red- and one blue- shifted local maximum for all observed transitions, where the red-shifted local maximum (``W44E1'') is located just behind the forward edge of molecular gas as delineated by OH masers. The blue-shifted local maximum (``W44E2'') is located further towards the interior of W44, where maser emission is also observed. In W44F, the thin filament of molecular gas ranging from the northwest to the southeast shows two local maxima in the red-shifted (``W44F2'' and ``W44F2'') and one in the blue-shifted velocity regime (``W44F3''), the latter in the very south of the filament, where the red-shifted emission is already very weak.
\item The individual spectra in the W44E1 and W44E2 exhibit two components: one narrow peak of ambient emission at $\sim$45 km s$^{-1}$, coincident with the emission in $^{13}$CO (3--2), and a broad component of shock emission. The shock-broadened lines are asymmetric but bear similar shapes in all $^{12}$CO transitions. Towards W44F the lower transitions are subject to strong absorption in their line centres. The lines are narrower and, except for W44F3, also more symmetric than in W44E.
\item The integrated intensities of CO (7--6) and (6--5) of the blue- and red lobe components in these position were compared with a large grid of shock models combined with radiative transfer modules. This comparison revealed non-stationary shock models to be compatible with the observations. These relatively young shocks have ages of no more than 3000 years, consistent with the remnant's age of 20,000 years. The pre-shock density for all models was found to be 10$^4$ cm$^{-3}$. The best-fitting models in W44E have velocities of $\sim$20 km s$^{-1}$, which seems low given the broad line widths. We therefore also presented the best-fitting models with velocities $\geq$30 km s$^{-1}$. The pre-shock magnetic field strengths of all best-fitting models in W44E are 100--200 $\mu$G, while in W44F we found values of 200 $\mu$G in all positions. 
\item To account for the integrated emission in CO (3--2) and (4--3), CO layers corresponding to post-shock gas had to be added to the shock models. The need for such additional layers stems from the one-dimensional nature of our modelling approach.
\item Based on the best-fitting shock models, we could estimate the shocked gas mass in one beam as well as the momentum- and energy injection per beam. In both regions, the shocked gas masses lie between $\sim$1 and 2 $M_{\odot}$, the momentum injection amounts to 1--5$\times$10$^6$~$M_{\odot}$ cm s$^{-1}$, and the energy injection to 1--8$\times$10$^{45}$ erg in one beam. This energy injection approximately corresponds to the fraction of the total initial energy that is expected to be dissipated in the observed area, assuming an isotropic explosion.

\item Our analysis can only be applied to shock components moving parallel to the line-of-sight, as revealed by the broad spectral line wings. Our results therefore correspond to mean properties of the shocked gas moving towards and away from us. 
Furthermore, we stress the present limitation of our analysis with respect to the very small number of unambiguously shock-tracing transitions we used in our fits. In the future, it will be necessary to base the modelling on additional shock-tracing transitions in order to eliminate modelling degeneracies.

\end{enumerate}

\begin{acknowledgements}
We are grateful to an anonymous referee for useful comments that helped to strengthen the paper. We thank Dale Frail for kindly providing us with the radio continuum map at 1442.5 MHz. S. Anderl acknowledges support by the DFG SFB 956, the International Max Planck Research School (IMPRS) for Astronomy and Astrophysics, and the Bonn-Cologne Graduate School of Physics and Astronomy. A. Gusdorf is grateful to Sylvie Cabrit for useful discussions. Moreover, he acknowledges support by the grant ANR-09-BLAN-0231-01 from the French {\it Agence Nationale de la Recherche} as part of the SCHISM project.
\end{acknowledgements}

\bibliographystyle{aa}
\bibliography{biblio}

\begin{appendix}

\section{CO Tables}

\begin{landscape}
 \begin{table*}
 \caption[]{Integrated intensities of rotational CO lines (given in K km s$^{-1}$) for our best fitting shock models in W44E, as listed in Table \ref{W44Eshock}. All models are CJ-type shocks at a pre-shock density of $n{}_{\rm H}$~=~10${}^4$ cm${}^{-3}$. In the table they are labeled as ($b$; velocity (km s$^{-1}$); age (yr)).}

\begin{center}
$$
\begin{tabular}{lllllllll}
\hline
           \noalign{\smallskip}

Transition &(2;35;915) & (1.75;20;1745) & (1.5,30,1190) & (1.75;20;2240) & (2;30;1390)& (1.5;22;1945) & (1.75;35;680) & (1;20;860)\\
           \noalign{\smallskip}
\hline
\hline
           \noalign{\smallskip}
CO (1--0) & 3.52e+00 & 7.30e+00 & 4.78e+00 & 8.86e+00 & 6.82e+00 & 7.12e+00 & 2.46e+00 & 3.29e+00\\ 
CO (2--1) & 1.31e+01 & 2.18e+01 & 1.80e+01 & 2.96e+01 & 2.42e+01 & 2.58e+01 & 9.13e+00 & 1.07e+01\\ 
CO (3--2) & 2.72e+01 & 3.69e+01 & 3.73e+01 & 5.40e+01 & 4.74e+01 & 5.05e+01 & 1.90e+01 & 1.96e+01\\ 
CO (4--3) & 3.62e+01 & 4.26e+01 & 5.10e+01 & 6.50e+01 & 6.12e+01 & 6.38e+01 & 2.54e+01 & 2.45e+01\\ 
CO (5--4) & 3.58e+01 & 3.92e+01 & 5.33e+01 & 6.09e+01 & 6.11e+01 & 6.18e+01 & 2.53e+01 & 2.41e+01\\ 
CO (6--5) & 2.99e+01 & 3.22e+01 & 4.80e+01 & 4.96e+01 & 5.33e+01 & 5.15e+01 & 2.12e+01 & 2.10e+01\\ 
CO (7--6) & 2.27e+01 & 2.52e+01 & 4.03e+01 & 3.76e+01 & 4.37e+01 & 3.96e+01 & 1.62e+01 & 1.74e+01\\ 
CO (8--7) & 1.65e+01 & 1.94e+01 & 3.33e+01 & 2.76e+01 & 3.53e+01 & 2.95e+01 & 1.19e+01 & 1.43e+01\\ 
CO (9--8) & 1.18e+01 & 1.47e+01 & 2.74e+01 & 2.00e+01 & 2.86e+01 & 2.16e+01 & 8.59e+00 & 1.15e+01\\ 
CO (10--9) & 8.33e+00 & 1.10e+01 & 2.27e+01 & 1.43e+01 & 2.33e+01 & 1.56e+01 & 6.14e+00 & 9.09e+00\\ 
CO (11--10) & 5.86e+00 & 7.92e+00 & 1.86e+01 & 9.98e+00 & 1.88e+01 & 1.11e+01 & 4.36e+00 & 6.92e+00\\ 
CO (12--11) & 4.10e+00 & 5.46e+00 & 1.51e+01 & 6.74e+00 & 1.49e+01 & 7.69e+00 & 3.08e+00 & 5.04e+00\\ 
CO (13--12) & 2.86e+00 & 3.57e+00 & 1.17e+01 & 4.40e+00 & 1.13e+01 & 5.14e+00 & 2.17e+00 & 3.48e+00\\ 
CO (14--13) & 2.03e+00 & 2.25e+00 & 8.59e+00 & 2.81e+00 & 8.04e+00 & 3.37e+00 & 1.55e+00 & 2.33e+00\\ 
CO (15--14) & 1.43e+00 & 1.37e+00 & 5.68e+00 & 1.76e+00 & 5.09e+00 & 2.16e+00 & 1.10e+00 & 1.50e+00\\ 
CO (16--15) & 1.03e+00 & 8.83e-01 & 3.34e+00 & 1.15e+00 & 2.85e+00 & 1.43e+00 & 7.92e-01 & 9.98e-01\\ 
CO (17--16) & 7.37e-01 & 5.71e-01 & 1.56e+00 & 7.62e-01 & 1.25e+00 & 9.55e-01 & 5.70e-01 & 6.62e-01\\ 
CO (18--17) & 5.40e-01 & 3.77e-01 & 1.05e+00 & 5.08e-01 & 8.27e-01 & 6.44e-01 & 4.21e-01 & 4.51e-01\\ 
CO (19--18) & 3.96e-01 & 2.51e-01 & 7.19e-01 & 3.41e-01 & 5.56e-01 & 4.37e-01 & 3.12e-01 & 3.10e-01\\ 
CO (20--19) & 2.93e-01 & 1.69e-01 & 5.02e-01 & 2.31e-01 & 3.80e-01 & 2.99e-01 & 2.33e-01 & 2.15e-01\\ 
CO (21--20) & 2.15e-01 & 1.14e-01 & 3.51e-01 & 1.57e-01 & 2.61e-01 & 2.04e-01 & 1.72e-01 & 1.49e-01\\ 
CO (22--21) & 1.58e-01 & 7.76e-02 & 2.48e-01 & 1.07e-01 & 1.81e-01 & 1.40e-01 & 1.28e-01 & 1.04e-01\\ 
CO (23--22) & 1.16e-01 & 5.31e-02 & 1.76e-01 & 7.35e-02 & 1.26e-01 & 9.61e-02 & 9.46e-02 & 7.30e-02\\ 
CO (24--23) & 8.50e-02 & 3.66e-02 & 1.26e-01 & 5.08e-02 & 8.82e-02 & 6.63e-02 & 7.01e-02 & 5.15e-02\\ 
CO (25--24) & 6.23e-02 & 2.54e-02 & 9.05e-02 & 3.53e-02 & 6.19e-02 & 4.60e-02 & 5.19e-02 & 3.64e-02\\ 
CO (26--25) & 4.56e-02 & 1.77e-02 & 6.51e-02 & 2.47e-02 & 4.36e-02 & 3.20e-02 & 3.83e-02 & 2.59e-02\\ 
CO (27--26) & 3.33e-02 & 1.25e-02 & 4.69e-02 & 1.74e-02 & 3.08e-02 & 2.24e-02 & 2.83e-02 & 1.85e-02\\ 
CO (28--27) & 2.42e-02 & 8.82e-03 & 3.37e-02 & 1.24e-02 & 2.17e-02 & 1.57e-02 & 2.08e-02 & 1.33e-02\\ 
CO (29--28) & 1.76e-02 & 6.28e-03 & 2.43e-02 & 8.83e-03 & 1.53e-02 & 1.10e-02 & 1.52e-02 & 9.55e-03\\ 
CO (30--29) & 1.27e-02 & 4.48e-03 & 1.75e-02 & 6.32e-03 & 1.08e-02 & 7.79e-03 & 1.11e-02 & 6.87e-03\\ 
CO (31--30) & 9.12e-03 & 3.22e-03 & 1.25e-02 & 4.55e-03 & 7.62e-03 & 5.51e-03 & 8.06e-03 & 4.95e-03\\ 
CO (32--31) & 6.50e-03 & 2.31e-03 & 8.93e-03 & 3.28e-03 & 5.35e-03 & 3.91e-03 & 5.81e-03 & 3.57e-03\\ 
CO (33--32) & 4.60e-03 & 1.66e-03 & 6.33e-03 & 2.36e-03 & 3.74e-03 & 2.77e-03 & 4.15e-03 & 2.56e-03\\ 
CO (34--33) & 3.22e-03 & 1.18e-03 & 4.45e-03 & 1.69e-03 & 2.59e-03 & 1.95e-03 & 2.92e-03 & 1.83e-03\\ 
CO (35--34) & 2.21e-03 & 8.37e-04 & 3.08e-03 & 1.19e-03 & 1.78e-03 & 1.36e-03 & 2.03e-03 & 1.29e-03\\ 
CO (36--35) & 1.49e-03 & 5.81e-04 & 2.09e-03 & 8.29e-04 & 1.19e-03 & 9.31e-04 & 1.38e-03 & 8.92e-04\\ 
CO (37--36) & 9.64e-04 & 3.91e-04 & 1.37e-03 & 5.58e-04 & 7.75e-04 & 6.19e-04 & 9.00e-04 & 5.98e-04\\ 
CO (38--37) & 5.90e-04 & 2.49e-04 & 8.47e-04 & 3.55e-04 & 4.78e-04 & 3.90e-04 & 5.55e-04 & 3.79e-04\\ 
CO (39--38) & 3.24e-04 & 1.42e-04 & 4.71e-04 & 2.02e-04 & 2.64e-04 & 2.20e-04 & 3.07e-04 & 2.15e-04\\ 
CO (40--39) & 1.34e-04 & 6.11e-05 & 1.98e-04 & 8.72e-05 & 1.11e-04 & 9.42e-05 & 1.28e-04 & 9.25e-05\\ 
\hline
           \noalign{\smallskip}
\end{tabular}
$$
\end{center}
\label{tbA1}
\end{table*}
\end{landscape}

%CO W44F

 \begin{table*}
 \caption[]{Integrated intensities of rotational CO lines (given in K km s$^{-1}$) for our best fitting shock models in W44F, as listed in Table \ref{W44Fshock}. All models are CJ-type shocks at a pre-shock density of $n{}_{\rm H}$~=~10${}^4$ cm${}^{-3}$. In the table they are labeled as ($b$; velocity (km s$^{-1}$); age (yr)).}

\begin{center}
$$
\begin{tabular}{llllllllll}
\hline
           \noalign{\smallskip}

Transition & Freq (GHz) & E$_{\rm up}$ (K) & (2;20;2900) & (2;22;2965) & (2;25;2380)\\
           \noalign{\smallskip}
\hline
\hline
           \noalign{\smallskip}
CO (1--0) & 115.27 & 5.5300 & 1.17e+01 & 1.21e+01 & 1.00e+01 \\ 
CO (2--1) & 230.54 & 16.600 & 3.98e+01 & 4.42e+01 & 3.69e+01 \\ 
CO (3--2) & 345.80 & 33.190 & 7.27e+01 & 8.49e+01 & 7.26e+01 \\ 
CO (4--3) & 461.04 & 55.320 & 8.73e+01 & 1.05e+02 & 9.18e+01 \\ 
CO (5--4) & 576.27 & 82.970 & 8.14e+01 & 1.00e+02 & 8.87e+01 \\ 
CO (6--5) & 691.47 & 116.16 & 6.56e+01 & 8.15e+01 & 7.34e+01 \\ 
CO (7--6) & 806.65 & 154.87 & 4.88e+01 & 6.08e+01 & 5.57e+01 \\ 
CO (8--7) & 921.80 & 199.11 & 3.50e+01 & 4.33e+01 & 4.06e+01 \\ 
CO (9--8) & 1036.9 & 248.88 & 2.47e+01 & 3.02e+01 & 2.91e+01 \\ 
CO (10--9) & 1152.0 & 304.16 & 1.73e+01 & 2.08e+01 & 2.07e+01 \\ 
CO (11--10) & 1267.0 & 364.97 & 1.18e+01 & 1.41e+01 & 1.45e+01 \\ 
CO (12--11) & 1382.0 & 431.29 & 7.89e+00 & 9.42e+00 & 9.97e+00 \\ 
CO (13--12) & 1496.9 & 503.13 & 5.09e+00 & 6.10e+00 & 6.63e+00 \\ 
CO (14--13) & 1611.8 & 580.49 & 3.22e+00 & 3.90e+00 & 4.31e+00 \\ 
CO (15--14) & 1726.6 & 663.35 & 1.99e+00 & 2.45e+00 & 2.73e+00 \\ 
CO (16--15) & 1841.3 & 751.72 & 1.29e+00 & 1.60e+00 & 1.80e+00 \\ 
CO (17--16) & 1956.0 & 845.59 & 8.46e-01 & 1.06e+00 & 1.18e+00 \\ 
CO (18--17) & 2070.6 & 944.97 & 5.59e-01 & 7.01e-01 & 7.92e-01 \\ 
CO (19--18) & 2185.1 & 1049.8 & 3.72e-01 & 4.68e-01 & 5.32e-01 \\ 
CO (20--19) & 2299.6 & 1160.2 & 2.51e-01 & 3.16e-01 & 3.60e-01 \\ 
CO (21--20) & 2413.9 & 1276.1 & 1.69e-01 & 2.13e-01 & 2.43e-01 \\ 
CO (22--21) & 2528.2 & 1397.4 & 1.15e-01 & 1.44e-01 & 1.64e-01 \\ 
CO (23--22) & 2642.3 & 1524.2 & 7.86e-02 & 9.81e-02 & 1.12e-01 \\ 
CO (24--23) & 2756.4 & 1656.5 & 5.42e-02 & 6.72e-02 & 7.59e-02 \\ 
CO (25--24) & 2870.3 & 1794.2 & 3.78e-02 & 4.64e-02 & 5.18e-02 \\ 
CO (26--25) & 2984.2 & 1937.4 & 2.65e-02 & 3.23e-02 & 3.54e-02 \\ 
CO (27--26) & 3097.9 & 2086.1 & 1.88e-02 & 2.26e-02 & 2.43e-02 \\ 
CO (28--27) & 3211.5 & 2240.2 & 1.34e-02 & 1.59e-02 & 1.67e-02 \\ 
CO (29--28) & 3325.0 & 2399.8 & 9.63e-03 & 1.13e-02 & 1.16e-02 \\ 
CO (30--29) & 3438.4 & 2564.8 & 6.94e-03 & 8.08e-03 & 8.02e-03 \\ 
CO (31--30) & 3551.6 & 2735.3 & 5.03e-03 & 5.79e-03 & 5.58e-03 \\ 
CO (32--31) & 3664.7 & 2911.1 & 3.65e-03 & 4.16e-03 & 3.89e-03 \\ 
CO (33--32) & 3777.6 & 3092.4 & 2.64e-03 & 2.98e-03 & 2.71e-03 \\ 
CO (34--33) & 3890.4 & 3279.1 & 1.90e-03 & 2.13e-03 & 1.88e-03 \\ 
CO (35--34) & 4003.1 & 3471.3 & 1.35e-03 & 1.51e-03 & 1.30e-03 \\ 
CO (36--35) & 4115.6 & 3668.8 & 9.42e-04 & 1.05e-03 & 8.79e-04 \\ 
CO (37--36) & 4228.0 & 3871.7 & 6.36e-04 & 7.02e-04 & 5.79e-04 \\ 
CO (38--37) & 4340.1 & 4080.0 & 4.05e-04 & 4.46e-04 & 3.61e-04 \\ 
CO (39--38) & 4452.2 & 4293.6 & 2.32e-04 & 2.54e-04 & 2.03e-04 \\ 
CO (40--39) & 4564.0 & 4512.7 & 9.99e-05 & 1.09e-04 & 8.64e-05 \\ 
 \hline
           \noalign{\smallskip}
\end{tabular}
$$
\end{center}
\label{tbA5}
\end{table*}

\end{appendix}

\end{document}